\documentclass[11pt,prd,aps,floatfix,nofootinbib]{revtex4-1}

\usepackage{graphicx}
\usepackage{subfigure}

\begin{document}

%// Preprint Number
\vspace*{-10mm}
\begin{flushright}
\normalsize
UT-CCS-60        \\
\end{flushright}

%Title of paper
\title{
Electromagnetic form factor of pion from $N_f=2+1$ dynamical flavor QCD
}

\author{
   Oanh Hoang Nguyen$^{a}$, 
   Ken-Ichi Ishikawa$^{b}$, 
   Akira Ukawa$^{a,c}$, 
   Naoya Ukita$^{c}$\\ 
   for PACS-CS Collaboration
}

\affiliation{
   $^a$Graduate School of Pure and Applied Sciences, University of Tsukuba, Ibaraki 305-8571, Japan
   \\
   $^b$Department of Physics, Hiroshima University,  Higashi-Hiroshima, Hiroshima 739-8526, Japan
   \\
   $^c$Center for Computational Sciences, University of Tsukuba, Ibaraki 305-8577, Japan
}

\date{\today}

\begin{abstract}
We present a calculation of the electromagnetic form factor of the pion in $N_f=2+1$ flavor lattice 
QCD.  Calculations are made on the PACS-CS gauge field configurations generated using Iwasaki gauge action 
and Wilson-clover quark action on a $32^3\times64$ lattice volume with the lattice spacing estimated 
as $a=0.0907(13)$~fm at the physical point. Measurements of the form factor are made using the technique of partially twisted 
boundary condition to reach small momentum transfer as well as periodic boundary condition with 
integer momenta.  Additional improvements including random wall source techniques and a judicious choice of 
momenta carried by the incoming and outgoing quarks are employed for error reduction. 
Analyzing the form factor data for the pion mass at $M_\pi \approx 411$~MeV and 296~MeV, 
we find that the NNLO SU(2) chiral perturbation theory fit yields $\left< r^2\right>=0.441 \pm 0.046~{\rm fm}^2$  
for the pion charge radius at the physical pion mass.  Albeit the error is quite large, this is consistent with   
the experimental value of $0.452\pm 0.011~{\rm fm}^2$.  Below $M_\pi\approx 300$~MeV, we 
find that statistical fluctuations in the pion two- and three-point functions become too large to 
extract statistically meaningful averages on a $32^3$ spatial volume.  We carry out a sample calculation on a $64^4$ lattice 
with the quark masses close to the physical point, which suggests that form factor calculations at the 
physical point become feasible by enlarging lattice sizes to $M_\pi L\approx 4$.
\end{abstract}


\maketitle

\clearpage

\section{Introduction}

The electromagnetic form factor of pion is an interesting quantity to investigate in lattice QCD.
Experimentally it has been measured in a set of experiments \cite{pdg}. 
Together with the nucleon form factors, it provides the first 
important test case of our understanding of hadron structure that arises from the quark content.  
From lattice QCD point of view, form factor calculations represent one of the first steps going beyond 
static quantities like the mass spectrum which require only two-point functions.  
The pion form factor is a natural first choice in this direction 
since usually pion Green's functions are statistically the most stable quantities in lattice QCD measurements. 
An interesting point with the pion form factor $G_\pi(q^2)$ is its slope at the origin as a function of the momentum transfer squared $q^2$, 
{\it i.e.,} the pion charge radius defined by 
\begin{equation}
\langle r^2\rangle =6\frac{dG_\pi(q^2)}{dq^2}\vert_{q^2=0}.
\end{equation}
It has been known for some time from chiral perturbation theory analysis \cite{gasserleutwyler,gasserleutwyler_2} 
that this quantity diverges logarithmically with vanishing 
pion mass squared. Quantitative confirmation of such a behavior would provide an important check on the control of chiral behavior 
in lattice QCD simulations toward the physical point.   

The pioneering lattice QCD calculations of the pion form factor appeared more than 20 years ago \cite{martinelli, draper}, and a 
number of studies were carried out over the years. Recently, with the development of simulations with dynamical quarks,  
several groups have attempted calculations with $N_f=2$ \cite{brommel, ETMC, JLQCD} and $N_f=2+1$ \cite{RBC} dynamical flavors 
using a variety of quark actions.  The $N_f=2$ calculations employed Wilson-clover \cite{brommel}, twisted mass \cite{ETMC} or 
overlap \cite{JLQCD} quark action, and explored the pion mass region down to $M_\pi\approx 300$~MeV.  The values for 
$\langle r^2\rangle$ from those simulations are significantly smaller than the experimental value, and NNLO fits of SU(2) chiral 
perturbation theory were needed to find consistency with it at the physical pion mass. 
For $N_f=2+1$ dynamical flavors, there has been a single calculation employing domain-wall quark action \cite{RBC}, which   
made measurements at a single pion mass of $M_\pi\approx 300$~MeV.  
Carrying out NLO analyses in SU(2) and SU(3) chiral perturbation theory, this work found $\langle r^2\rangle$ to be consistent with 
the experimental value at the upper edge of a 10\% error band.  

In this paper we present our calculation of the electromagnetic form factor of pion in $N_f=2+1$ dynamical flavor QCD using the Wilson-clover 
quark action.  For measurements we employ the $N_f=2+1$ PACS-CS gauge configurations generated on a $32^3\times 64$ lattice using 
the Iwasaki gauge action and the Wilson-clover action at a lattice spacing estimated to be $a=0.0907(13)$~fm at the physical point \cite{pacscs}. 
Since the pion mass on this gauge configuration set covers the range from $M_\pi\approx 700$MeV down to $156$~MeV, we are able to examine both 
the known range above $M_\pi\approx 300$MeV and a novel range below toward the physical pion mass.  

The paper is organized as follows. In Sec.~2 we present our method to calculate the pion form factor.  In order to access the region 
of small momentum transfer, we use the method of partially twisted boundary condition \cite{boyle2007,sachrajda2005,bedaque2005,jian}, 
and in order to fight increasing computational cost for smaller pion mass, we apply the method of random wall 
source \cite{Z2_1,Z2_2,Z2_3,Z2_4,Z2_5,RBC}.  
In addition we make use of a judicious choice of momenta carried by the 
incoming and outgoing quarks off the electromagnetic vertex, which helps in reducing statistical fluctuations in the form factor measurements. 
In Sec.~3 we present the results of pion form factor measurements, and in Sec.~4 
we analyze the data as a function of the momentum transfer squared and pion mass.  In particular we examine consistency with the predictions of 
chiral perturbation theory.  Finally, in Sec.~5,  we discuss our findings closer to the physical point including the results of our test 
calculation on a $64^4$ lattice with the quark masses tuned to the neighbour of the physical point. 
We end this work with conclusions in Sec.~6.  A preliminary report of this work was presented in \cite{OanhNguyen}.

\section{Methods}

\subsection{Pion electromagnetic form factor}

The electromagnetic pion form factor $G_\pi(Q^2)$ is defined by 
\begin{equation}
\label{eq:fund4}
\left< \pi^+(\vec{p'})|J_\mu|\pi^+(\vec{p})\right> = (p_\mu + p'_\mu)G_\pi(Q^2),
\end{equation}
where $Q^2=-q^2=-(p'-p)^2$ is the four-momentum transfer, and $J_\mu$ is the electromagnetic current given in $N_f=2+1$ QCD by 
\begin{equation}
\label{eq:fund3}
J_\mu = \frac{2}{3} \bar{u}\gamma_\mu{u}-\frac{1}{3} \bar{d}\gamma_\mu{d}-\frac{1}{3} \bar{s}\gamma_\mu{s}.
\end{equation}
In the limit of vanishing four-momentum transfer $Q^2=0$, the form factor equals unity due to the charge conservation. 

We extract the form factor from a suitable ratio of the pion two- and three-point functions. 
We use a ratio, which has the advantage of simultaneously reducing fluctuations and renormalizing the
current, defined as
\begin{equation}
\label{eq:ratio}
R(\tau)=\frac{C^{3pt}(\vec{p'},t_f;\vec{p},0;\tau)}{C^{3pt}(\vec{p'},t_f;\vec{p'},0;\tau)} \frac{C^{2pt}(\vec{p'},\tau)}{C^{2pt}(\vec{p},\tau)}
\times \frac{2E_\pi(\vec{p'})}{E_\pi(\vec{p})+E_\pi(\vec{p'})},
\end{equation}
which converges as
\begin{equation}
R(\tau)\to \frac{G^{\rm bare}_\pi(Q^2)}{G^{\rm bare}_\pi(0)}=G_\pi(Q^2),
\end{equation}
for large $\tau$ and $t_f$. $E_\pi(\vec{p})$ denotes pion energy for spatial momentum $\vec{p}$, 
$C^{3pt}(\vec{p'},t_f; \vec{p},0;\tau)$ is the three-point function with momenta $\vec{p}$ at the source
and $\vec{p'}$ at the sink $t_f$,
\begin{equation}
\label{eq:C3pt}
C^{3pt}(\vec{p'},t_f; \vec{p},0;\tau) = \left< \pi^+(\vec{p'},t_f) J_\mu(\tau) \pi^+(\vec{p},0)\right>,
\end{equation}
and $C^{2pt}(\vec{p},\tau)$ is the two-point function,
\begin{equation}
\label{eq:C2pt}
C^{2pt}(\vec{p},\tau) = \left< \pi^+(\vec{p},\tau) \pi^+(\vec{p},0)\right>.
\end{equation}
After contraction of quark fields, the three-point function consists of the connected and disconnected contributions.  
The latter vanishes after the gauge field average due to charge conjugation invariance, and hence need not be calculated.  Since we assume 
degeneracy of up and down quark masses in the present calculation, the connected contribution is equal to
\begin{equation}
\label{eq:3pt}
C^{3pt}(\vec{p'},t_f; \vec{p},0;\tau) = \sum_{\vec{y},\vec{x}}e^{-i\vec{p'}\vec{y}+i\vec{q}\vec{x}} 
\left<Tr[\gamma_5 D^{-1}(0,x) \gamma_\mu D^{-1}(x,y) \gamma_5 D^{-1}(y,0)]\right>.
\end{equation}
This contribution can be calculated by the traditional source method \cite{martinelli,draper}.

\subsection{Choice of momenta carried by quarks}

The ratio (\ref{eq:ratio}) makes use of two- and three-point functions in an appropriate combination 
to extract the form factor for the renormalized current. The presence of ratios guarantee that statistical fluctuations are 
suppressed.  Nonetheless, making simple choices such as $\vec{p}\ne 0$ and $\vec{p'}=0$, we have observed  
an increasingly larger fluctuation of the ratio as pion mass is reduced, 
and this trend worsens for larger momenta.  With an interesting choice of momenta, $\vec{p'}\ne \vec{p}$ but $|\vec{p'}|=|\vec{p}|$, 
the  ratio (\ref{eq:ratio}) simplifies to 
\begin{equation}
\label{eq:ratio3ptOnly}
R'(\tau)=\frac{C^{3pt}(\vec{p'},t_f;\vec{p},0;\tau)}{C^{3pt}(\vec{p'},t_f;\vec{p'},0;\tau)}.  
\end{equation}
Since the two-point functions as well as the ratio of energies drop out, leaving just the ratio of three-point functions, 
we expect this choice to yield better signals than those choices for which all factors are present. 
Furthermore, one can choose 6 permutations in momentum directions while keeping $|\vec{p'}|=|\vec{p}|$, gaining more statistics.

\subsection{Partially twisted boundary condition}

The minimum non-zero quark momentum $2\pi/La$ for the periodic boundary condition on a $32^3\times 64$ lattice 
with a 2~GeV inverse lattice spacing is about 0.4~GeV.
To probe the region of smaller momentum transfer as well as to improve the resolution of four-momentum transfer,
we apply the method of partially twisted boundary condition \cite{boyle2007,sachrajda2005,bedaque2005,jian} in which 
valence quark fields are subjected to twisted boundary condition while periodic boundary condition is kept for sea quark 
fields. If one imposes the boundary condition given by 
\begin{equation}
\label{eq:fund45}
\psi(x+Le_j)=e^{2\pi i\theta_j}\psi (x), \qquad j=1,2,3,
\end{equation}
on a valence quark field, the spatial momentum of that quark is quantized according to 
\begin{equation}
\label{eq:fund46}
p_j = \frac{2\pi n_j}{L} + \frac{2\pi \theta_j}{L}, \qquad j=1,2,3,
\end{equation}
where $L$ denotes the spatial lattice size, $e_j$ the unit vector in the spatial $j$-th direction and $\theta_j$ real parameter.
In this way one can explore arbitrarily small momentum on the lattice by adjusting the value of twist $\theta_j$. 

For the meson two-point function consisting of quark and anti-quark propagators, 
we apply the twist only to quark and not to antiquark or vice versa.  
Similarly, for the three-point function, we twist only one or two out of the three quark propagators.  
In other words, we pretend that each valence quark line in the two- and three-point function quark diagrams carry a different flavor
and select the appropriate flavor to apply twisting.  In this way we can avoid a twist of a quark line cancelled by the opposite 
twist of the antiquark line carrying the same flavor \cite{boyle2007}.  
This procedure and the twisting of only valence quarks mean that we deal with partially quenched QCD with a different flavor symmetry 
content in the valence and sea quark sectors. 
As was discussed in detail in \cite{sachrajda2005,bedaque2005} using chiral perturbation theory, the associated effects are expected to 
appear as finite-size effects exponentially small in spatial volume for channels which do not have final-state interactions such as 
three-point functions for form factor calculations.  Since terms of such magnitude are also present in unitary theory with periodic 
valence and sea quarks, we ignore this issue in the present work.  

The twisted boundary condition can be imposed on a periodic quark field configuration by the following transformation
\begin{equation}
\label{eq:fund47}
\psi(x) \longrightarrow U(\theta,x) \psi(x) = e^{2\pi i \sum^{3}_{j=1}{\theta_j x_j/L}} \psi(x).
\end{equation}
In practice we transfer the twist from the quark sector to the gluon sector by an ${\rm U(1)}$ transformation on the spatial gluon link 
fields given by
\begin{equation}
\label{eq:fund48}
U_i(x) \rightarrow U^\theta_i(x)=e^{2\pi i \theta_i/L}U_i(x), \qquad i=1,2,3.
\end{equation}
Thus valence quark propagators are solved with the periodic boundary condition but on the PACS-CS gauge configurations 
twisted by the U(1) transformation above.

\begin{figure}[t]
\centering
\subfigure[Pion effective energies.]{\includegraphics[scale=.5363]{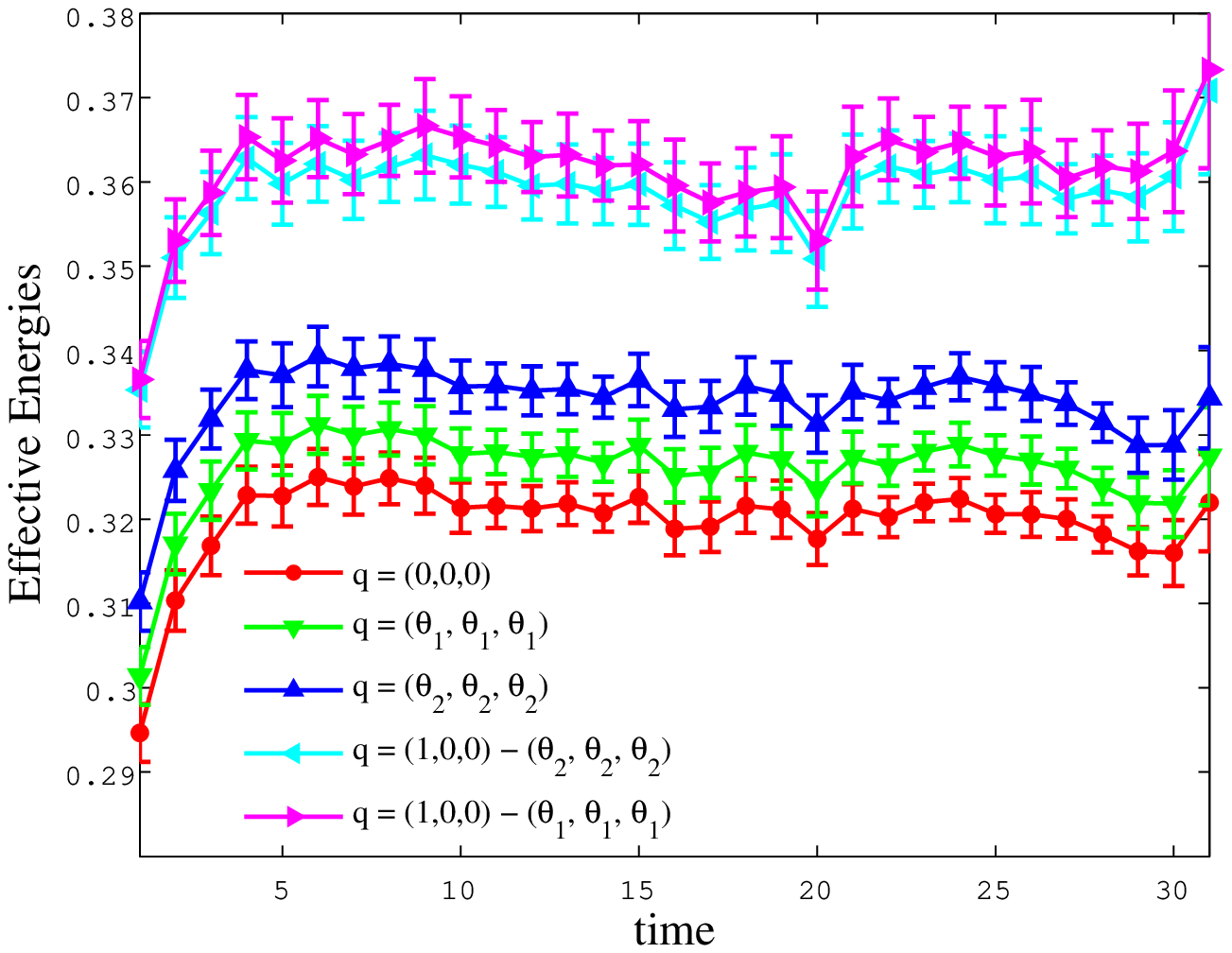}}
\subfigure[Energy momentum dispersion relation.]{\includegraphics[scale=.5363]{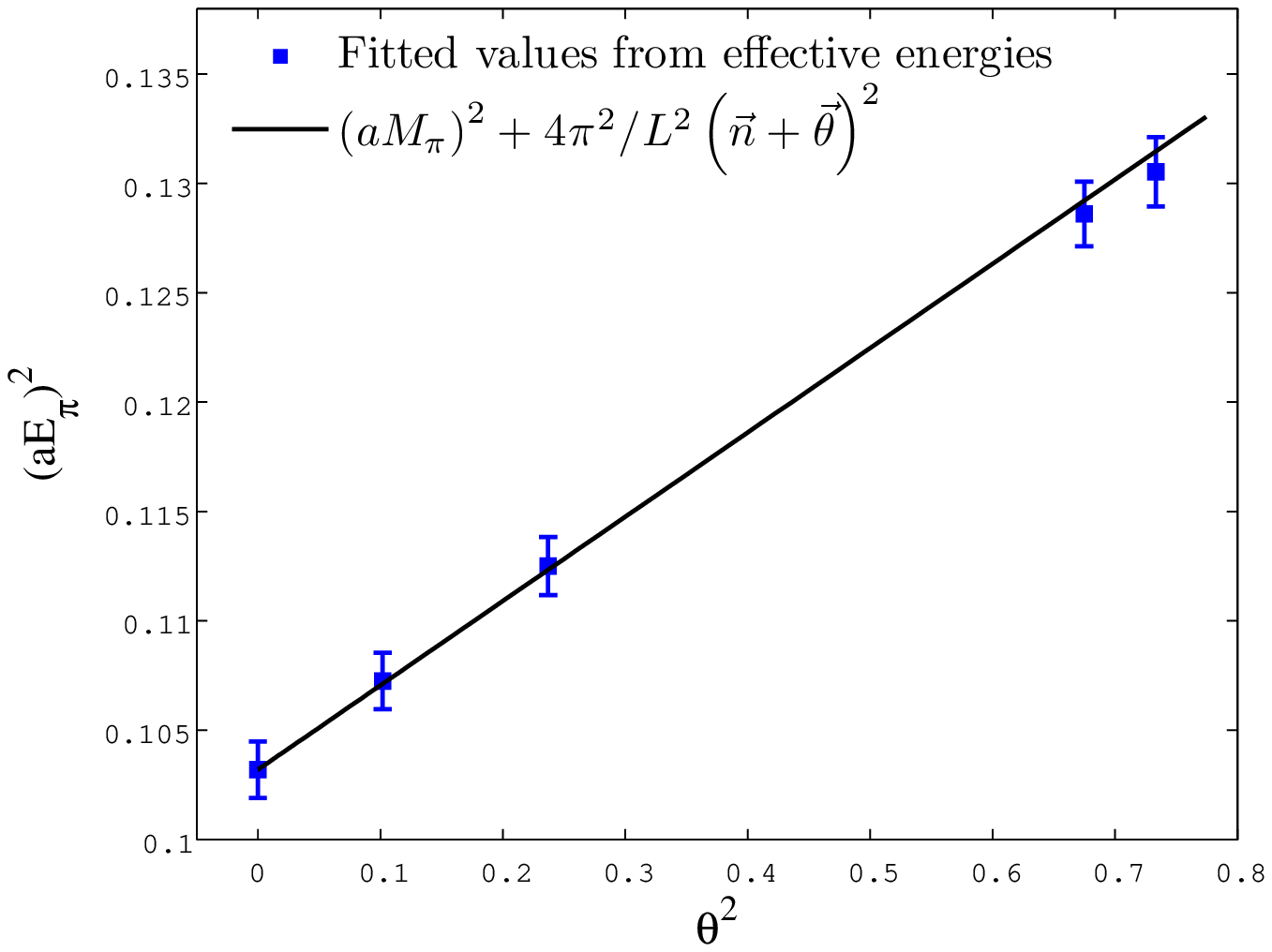}}
\caption{Check of validity of the twisted boundary condition at $\kappa_{s}=0.1364, \kappa_{ud}=0.13700$ where $M_\pi \approx 702$~MeV.  
Values of twist angle used in test are $\theta=0.1842$ and 0.2811. Measurements are made on 40 configurations.}
\label{fig:checkTBC13700}
\end{figure}

In order to check that the term $\frac{2\pi \theta_j}{L}$ acts as true physical momentum, we carried out a test of the energy-momentum 
dispersion relation of the pion on some PACS-CS configurations. Of the two valence quarks inside the pion, 
we twisted one quark with a twist angle $\vec{\theta}=(\theta, \theta, \theta)$ and left the other untwisted.
In Fig.~\ref{fig:checkTBC13700}(a) we plot the effective energy for the ground state, two values of the twist angle and their combination with the first 
integer momenta at the hopping parameters $\kappa_{s}=0.1364$, $\kappa_{ud}=0.13700$ where $M_\pi \approx 702$~MeV. The propagator is fitted 
over $t=7-27$ to extract the energy $E(\vec p)$. Errors are estimated by the jackknife method with the bin size of 100 trajectories.
The results are plotted in Fig.~\ref{fig:checkTBC13700}(b), together with the expected behavior,
\begin{equation}
\label{eq:DispersionRel}
E\left(\vec p \right)^2 = \left( aM_\pi \right)^2 + \left( \frac{2\pi}{L}\vec n + \frac{2\pi}{L}\vec\theta \right)^2, 
\end{equation}
which demonstrates clearly that the term $\frac{2\pi \theta_j}{L}$ acts as true physical momentum. 
The two data points on the right represent combinations of an integer momentum $(1,0,0)$ and a twist. 
The energy-momentum relation is correctly reproduced in this case as well. 

\subsection{Random wall source}

At light quark masses the computing cost for inversion of Dirac operator becomes very expensive. 
Thus we have employed some improvements for obtaining the form factor with acceptable statistical errors at reasonable computing time. 
The first improvement is to utilize the random wall source.  This method has a long history and has been applied to two-point functions in a 
variety of contexts.  More recently, applications to three-point functions have shown their effectiveness for form factor 
calculations \cite{ETMC,RBC}. We consider the use of $Z(2) \otimes Z(2)$ random noisy source as introduced in \cite{Z2_3}.

Consider a set of random sources whose real and imaginary components are randomly chosen from $Z(2)$ for each site, color and spin,
\begin{equation}
\label{eq:Impr2}
\{\eta^{(n)}(x)_{a\alpha} \in Z(2) \otimes Z(2) | n = 1...N\}.
\end{equation}
This set has the property that in the limit $N \rightarrow \infty$ 
\begin{equation}
\label{eq:Impr3}
\langle\eta_{a\alpha}^{(n)}(x) \eta_{b\beta}^{\dagger(n)}(y)\rangle_n = 
\frac{1}{N}\sum_{n=1}^N \eta_{a\alpha}^{(n)}(x) \eta_{b\beta}^{\dagger(n)}(y) \rightarrow \delta_{xy}\delta_{ab}\delta_{\alpha\beta}.
\end{equation}
To use this kind of source in calculating correlators, one can choose it to be a set of random wall source located at $t_0$,
\begin{eqnarray}
\label{eq:Impr4}
\eta_{a\alpha}^{(n)}(\vec{x},t|t_0) \in Z(2) \otimes Z(2) &|& t = t_0 \nonumber\\
= 0 &|& t \neq t_0,
\end{eqnarray}
\begin{equation}
\label{eq:Impr5}
\langle\eta_{a\alpha}^{(n)}(\vec{x},t|t_0) \eta_{b\beta}^{\dagger(n)}(\vec{y},t|t_0)\rangle_n = \delta_{xy}\delta_{ab}\delta_{\alpha\beta},
 \qquad N \rightarrow \infty.
\end{equation}
Making use of (\ref{eq:Impr5}) to rewrite the pion two-point function at zero momentum as,
\begin{eqnarray}
\label{eq:Impr6}
\nonumber
C(\tau;\vec{0}) &=& \sum_{\vec{x},\vec{y}} tr \left( D^{-1}(\vec{y},t; \vec{x},t_0) D^{-1\dagger}(\vec{y},t; \vec{x},t_0) \right) \\\nonumber
&=& \sum_{\vec{x},\vec{y},\vec{z}} 
\Big( D^{-1}_{a\alpha,b\beta}(\vec{y},t; \vec{x},t_0) \left[ \delta_{xz}\delta_{bc}\delta_{\beta\kappa} \right] 
D^{-1\dagger}_{c\kappa,a\alpha}(\vec{y},t; \vec{z},t_0) \Big)\\
&=& \sum_{\vec{y}} \left< \psi^{(n)}(\vec{y},t|t_0) \psi^{\dagger(n)}(\vec{y},t|t_0) \right>_n,
\end{eqnarray}
where $\psi^{(n)}$ is the solution vector of the Dirac equation,
\begin{equation}
\label{eq:Impr7}
\psi^{(n)}(\vec{y},t|t_0) = \sum_{\vec{x}} D^{-1}(\vec{y},t; \vec{x},t_0) \eta^{(n)}(\vec{x},t|t_0).
\end{equation}

\begin{figure}[t]
\centering
\subfigure[Effective pion mass with point source.]
{\includegraphics[scale=.625]{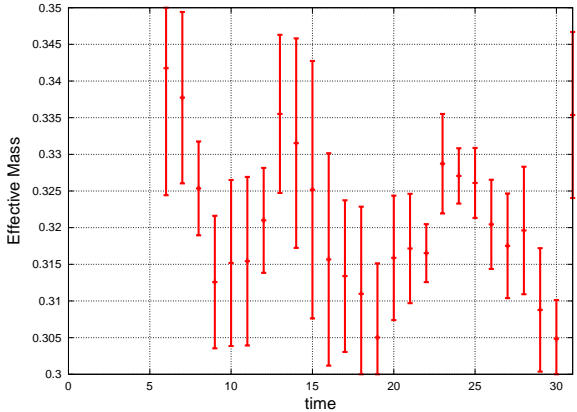}}
\subfigure[Effective pion mass with random $Z(2) \otimes Z(2)$ wall source.]
{\includegraphics[scale=.625]{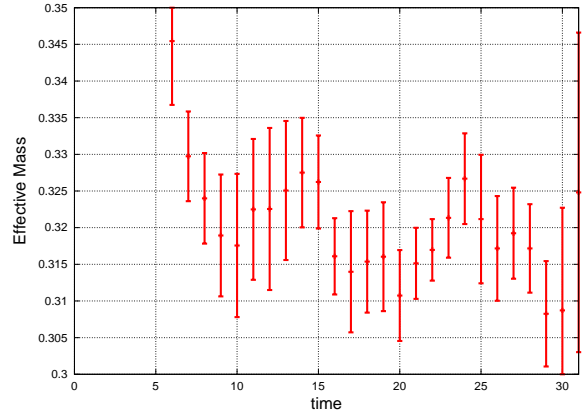}}\\
\subfigure[Effective pion mass with smeared source.]
{\includegraphics[scale=.625]{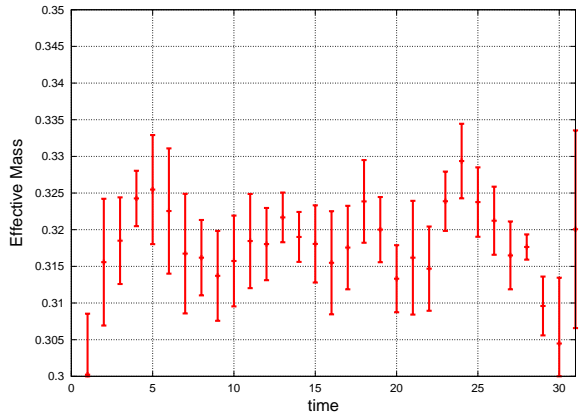}}
\caption{Comparison of pion effective masses calculated with (a) point source, (b) random $Z(2) \otimes Z(2)$ wall source, and (c) smeared source 
on a set of 10 configurations at $\kappa_s=0.1364,\kappa_{ud}=0.13700$ ($M_\pi \approx 702~$MeV).}
\label{fig:C2ptPointVSZ2}
\end{figure}

With a random $Z(2) \otimes Z(2)$ wall source the solution for quark propagator needs only single inversion instead of $3\times 4=12$ 
corresponding to color and Dirac components required 
for a point source or smeared source.
When the number of configurations in the ensemble is large enough, even if one uses a single random source for each configuration, 
(\ref{eq:Impr5}) is expected to hold in the ensemble average.  
One may expect to obtain meson correlators of a similar statistical quality as with the traditional point source with only 1/12 of
computing time. 

In Fig.~\ref{fig:C2ptPointVSZ2}, we compare the effective pion mass plot calculated 
on a set of 10 configurations at $\kappa_s=0.1364,\kappa_{ud}=0.13700$ ($M_\pi \approx 702~$MeV) from the PACS-CS ensemble 
using (a) point source, (b) single random $Z(2) \otimes Z(2)$ wall source, and (c) smeared source. 
We observe that the signal with the random $Z(2) \otimes Z(2)$ wall source is somewhat better than that for point source, while the signal 
for smeared source is better than that with the random $Z(2) \otimes Z(2)$ wall source.  Using 4 random wall sources for each configuration, 
we observed that the quality of signal becomes comparable to that for smeared source.  Since the computing time is still $4/12=1/3$ for 
the random wall source, we employ the method of random wall source with 4 sets of random wall sources in our measurements.  In addition 
we repeat measurements with the source located at $t=0,16,32,48$ since the time extent of our lattice is 64. 

\section{Measurements}

We apply our calculational setup to a subset of the PACS-CS gauge configurations \cite{pacscs} 
corresponding to the degenerate up-down hopping parameter
in the set $\kappa_{ud} = \{0.13700$, $0.13727$, $0.13754$, $0.13770\}$.
The hopping parameter of strange quark is fixed at $\kappa_s = 0.1364$.

The first set of measurements, which we call data set I, is made with an exponentially smeared source and local sink,  
setting the final pion at zero momentum $\vec{p'}=\vec{0}$ and varying that of the initial pion $\vec{p}$ in the three-point function. 
The fixed sink time $t_f$ in the ratio (\ref{eq:ratio}) needs to be chosen large enough to eliminate excited states contributions.
However, statistical fluctuations increase as $t_f$ increases, and examining measurement results, we choose $t_f=24$ to balance the 
two opposite features. 
The twist technique is applied to the quark running from the source to the current. 
Two values are chosen for the twist angle $\vec{\theta}=\left(\theta, \theta, \theta\right)$ such that 
the smallest four-momentum transfer of the current takes the value $Q^2({\rm GeV^2})=0.01841$ or $0.04237$.  
Adding integer momenta, we then collect data for $Q^2$ in the range $0.01841 {\rm ~GeV}^2 \leq Q^2 \leq 0.7302{\rm ~GeV}^2$.
The statistics of data set I is given in Table \ref{table:stat_dataset1} together with pion and kaon mass.  
Results of data set I have been previously reported in \cite{OanhNguyen}.

In order to extract the form factor, we fit the plateau of the ratio $R(\tau)$ by a constant.
The fitting range should be chosen around the symmetry point between the source and the sink, with additional 
considerations on the time interval required for the pion state to become dominant. 
Since we employ an exponential smeared source and a point sink, we shift the fitting range one time unit closer to the source 
than the symmetric point $t_f/2=12$. 

\begin{figure}[t]
\centering
\subfigure[Data set I: Ratio $R(\tau)$ defined in (\ref{eq:ratio}) at all 7 values of 4-momentum transfer.]
{\includegraphics[scale=.5363]{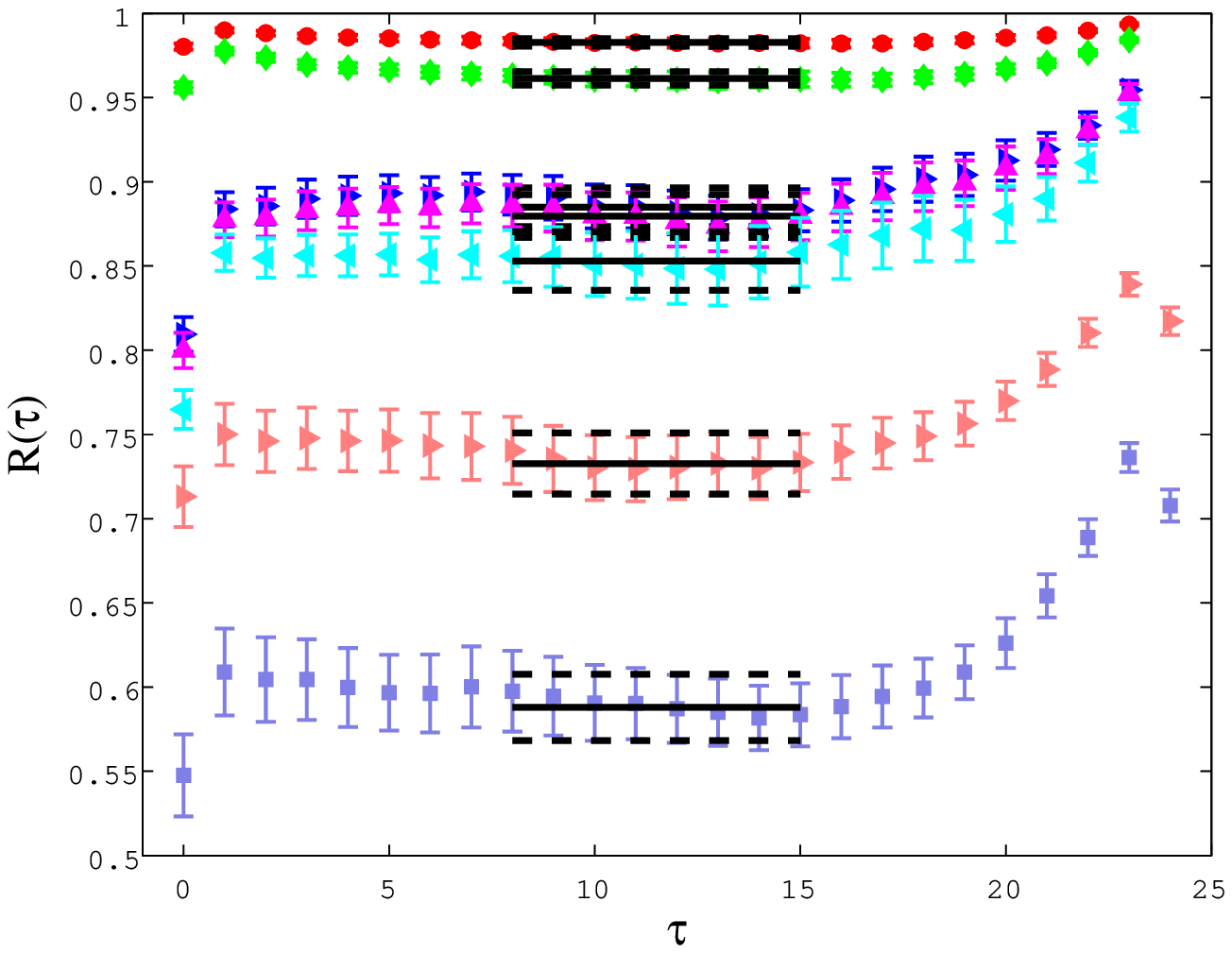}}
\subfigure[Data set II: Ratio $R'(\tau)$ defined in (\ref{eq:ratio3ptOnly}) at all 5 values of 4-momentum transfer.]
{\includegraphics[scale=.5363]{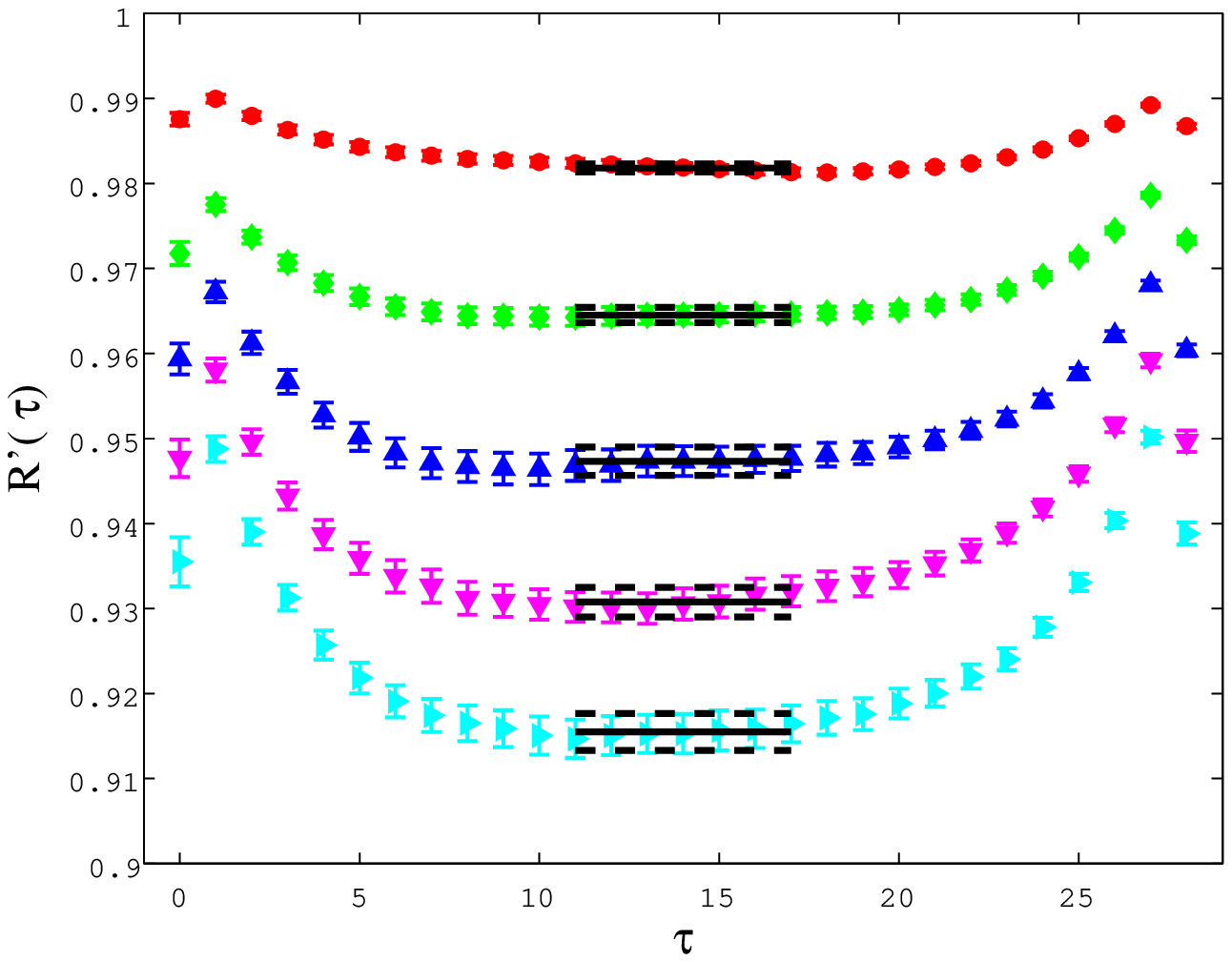}}
\caption{Ratios to extract the form factor as functions of the time slice $\tau$ of the current operator at $M_\pi \approx 702$~MeV.}
\label{fig:Plateau_PFF}
\end{figure}

\begin{figure}[t]
\centering
\subfigure[Data set I: Ratio $R(\tau)$ defined in (\ref{eq:ratio}) at 2 smallest 4-momentum transfers: $Q^2({\rm GeV^2})=0.01841,0.04237$.]
{\includegraphics[scale=.5363]{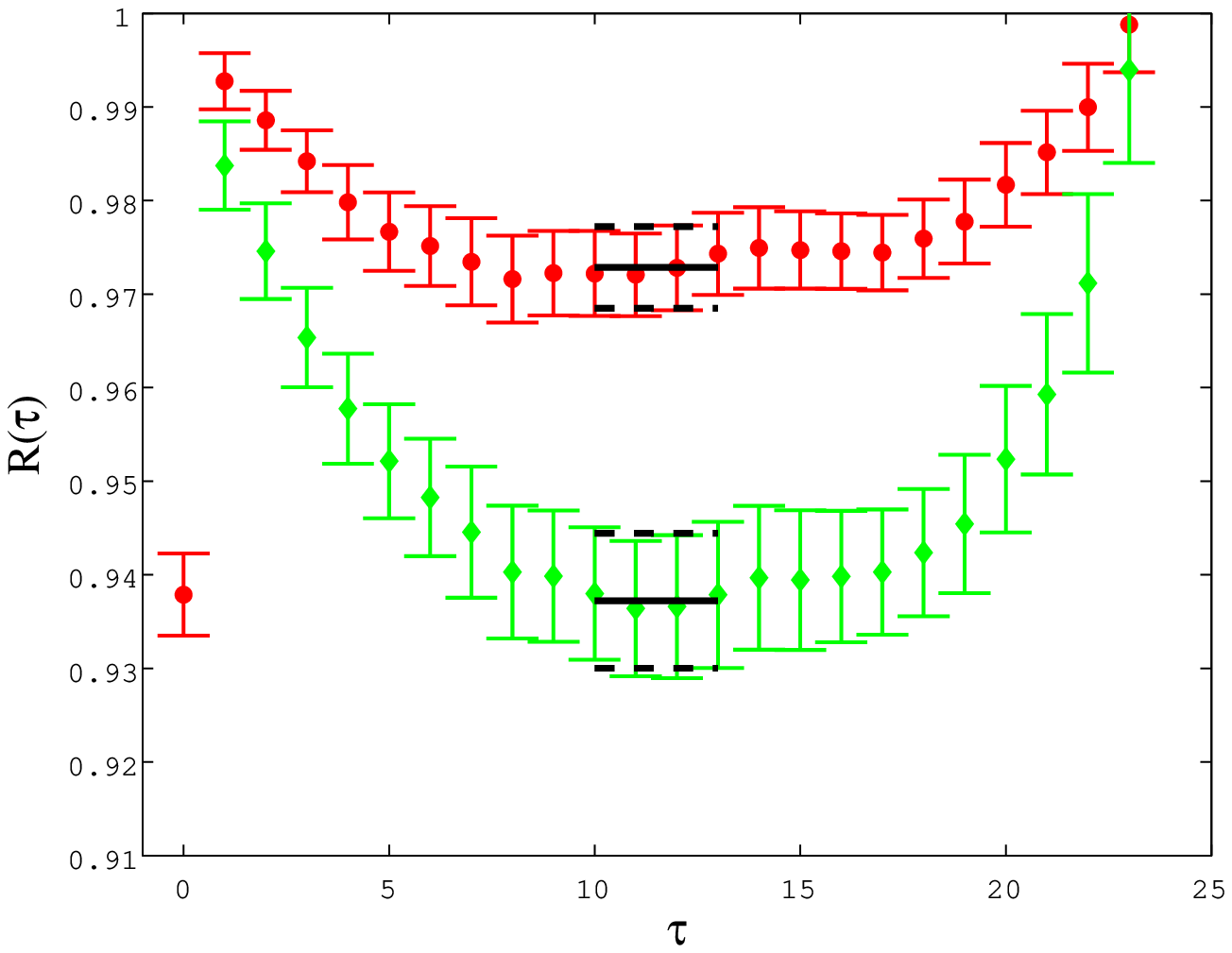}}
\subfigure[Data set II: Ratio $R'(\tau)$ defined in (\ref{eq:ratio3ptOnly}) at 4 smallest 4-momentum transfers: $Q^2({\rm GeV^2})=0.02,0.04,0.06,0.08$.]
{\includegraphics[scale=.5363]{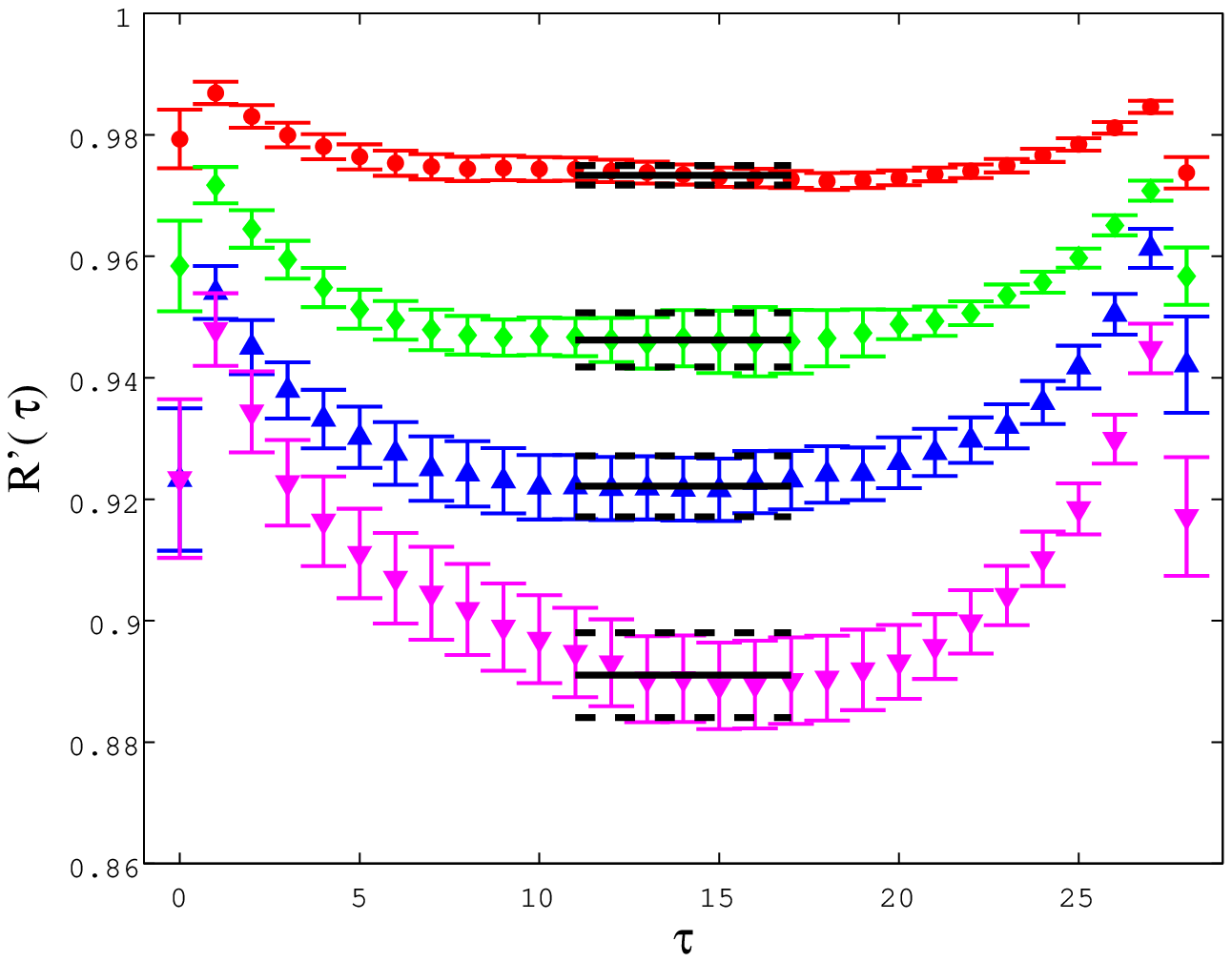}}
\caption{Ratios to extract the form factor as functions of the time slice $\tau$ of the current operator at $M_\pi \approx 296$~MeV.}
\label{fig:Plateau_PFF_296}
\end{figure}

In Fig.~\ref{fig:Plateau_PFF}(a) we plot the ratio $R(\tau)$ at various momentum transfer for the pion mass $M_\pi \approx 702$~MeV.
At this pion mass we have good signals for all 7 values of the four-momentum transfer.
There is a good plateau from $\tau=8$ to 15 for every momentum transfer.
Thus at this pion mass we can choose the fitting range from $\tau=8$ to 15 to extract the form factor.
However, as the pion mass decreases, the plateau signal becomes worse as exhibited in Fig.~\ref{fig:Plateau_PFF_296}(a) for the lightest case of $M_\pi\approx 296$~MeV 
where the ratio $R(\tau)$ at the two smallest momentum transfers, $Q^2({\rm GeV^2})=0.01841$ and $0.04237$, is shown. 
We then choose larger values for the starting point of the fitting range for better suppression of excited states at lighter pion masses. 
The error is estimated by the jackknife method using 10 configurations corresponding to 50 hybrid molecular dynamics time units 
as the bin size after checking saturation of the magnitude of error as function of bin size. 
Fit results for the pion form factor are listed in Table \ref{table:PFF_dataset1}. 

\begin{table}[tb]
\centering
\begin{tabular}{c c c c c c}
\hline
\hline
$\kappa_{ud}$ & $\kappa_{s}$ & $M_\pi$ (MeV) & $M_K$ (MeV) & \#conf measured & $\theta$ \\
\hline
  0.13700 & 0.1364 & 702 & 792 & 40 & 0.18423, 0.28112\\
  0.13727 & 0.1364 & 570 & 716 & 40 & 0.18467, 0.28265\\
  0.13754 & 0.1364 & 411 & 637 & 40 & 0.18585, 0.28672\\
  0.13770 & 0.1364 & 296 & 596 & 160 & 0.18814, 0.29450\\
\hline
\hline
\end{tabular}
\caption{Statistics of data set I.}
\label{table:stat_dataset1}
\end{table}

\begin{table}[!ht]
\centering
\begin{tabular}{c c c c c c c c c c}
\hline
\hline
\multicolumn{8}{c}
{$M_\pi \approx 702$~MeV, fit range:[8-15], bin size: $50\tau$}\\
\hline
$Q^2$(GeV$^2$) & 0.01842   & 0.04237   & 0.1163     & 0.1258     & 0.1682     & 0.3651     & 0.7302\\
$G_\pi(Q^2)$   & .9825(24) & .9609(43) & .8834(120) & .8780(134) & .8511(188) & .7313(186) & .5875(200)\\
\hline
\\
\hline
\multicolumn{8}{c}
{$M_\pi \approx 570$~MeV, fit range:[9-14], bin size: $50\tau$}\\
\hline
$Q^2$(GeV$^2$) & 0.01842   & 0.04237   & 0.1132     & 0.1223     & 0.1623     & 0.3651     & 0.7302\\
$G_\pi(Q^2)$   & .9836(37) & .9604(61) & .8816(154) & .8746(160) & .8400(184) & .6934(212) & .5191(189)\\
\hline
\\
\hline
\multicolumn{8}{c}
{$M_\pi \approx 411$~MeV, fit range:[9-14], bin size: $50\tau$}\\
\hline
 $Q^2$(GeV$^2$)  & 0.01841    & 0.04237    & 0.1062      & 0.1143      & 0.1495      & 0.3651      & 0.7302 \\
 $G_\pi(Q^2)$    & .9730(54)  & .9428(66)  & .9036(315)  & .8920(319)  & .8805(476)  & .5999(535)  & .4706(574)\\
\hline
\\
\hline
\multicolumn{8}{c}
{$M_\pi \approx 296$~MeV, fit range:[10-13], bin size: $50\tau$,} \\
\hline
 $Q^2$(GeV$^2$) & 0.01842    & 0.04237   & 0.09612    & 0.1030     & 0.1324     & 0.3651      & 0.7302\\
 $G_\pi(Q^2)$   & .9728(44)  & .9372(72) & .8624(310) & .8456(343) & .7929(452) & .9758(3376) & .6115(1943)\\
\hline
\hline
\end{tabular}
\caption{Pion form factor obtained with data set I.}
\label{table:PFF_dataset1}
\end{table}

We observe in Table \ref{table:PFF_dataset1} for data set I that the error for the form factor becomes large 
toward small pion mass and large momentum transfer. 
In order to improve the quality of data, we repeat the measurements (i) choosing the incoming and outgoing pions 
to have momenta with the same magnitude $|\vec{p'}|=|\vec{p}|$, and (ii) applying 4 random $Z(2) \otimes Z(2)$ wall 
sources located at $t=0,16,32,48$ for the lattice time extent of 64.
The twist technique is applied to two quarks running from the source to the current and from the current to the sink.
Five values are employed for the twist angle of form $\vec{\theta}=\left(\theta, 0, 0\right)$ and its permutations such that 
four-momentum transfer of the current takes the value $Q^2({\rm GeV}^2)=0.02, 0.04, 0.06, 0.08, 0.10$. 
Those values of $\theta$ are independent of $M_\pi$ as is easily checked for the momentum configuration chosen here.
The fixed sink time $t_f$ is chosen to be 28, larger than that of data set I, 
for better suppression of excited states and also from examination of the dependence of the ratio on $t_f$.
The fitting range is chosen symmetric around $t=14$ since the source is local after averaging over the $Z(2) \otimes Z(2)$ random numbers.
We call this set of data as data set II. 
Statistics and results of data set II are tabulated in Table \ref{table:stat_dataset2} and Table \ref{table:PFF_DataSet2}.
Results for $R'(\tau)$ for $M_\pi \approx 702$~MeV are plotted in Fig.~\ref{fig:Plateau_PFF}(b). 
One can see that the form factors of the data set II have much smaller error bars compared to those of data set I. 
We also plot results for the case of pion mass $M_\pi \approx 296 \rm MeV$ in Fig. \ref{fig:Plateau_PFF_296}(b). 

\begin{table}[!ht]
\centering
\begin{tabular}{c c c c c c c}
\hline
\hline
$\kappa_{ud}$ & $\kappa_{s}$ & $M_\pi$ (MeV) & $M_K$ (MeV) & \#conf measured & $Q^2($ GeV $^2)$ \\
\hline
0.13700 & 0.1364 & 702 & 792 & 40 & 0.02, 0.04, 0.06, 0.08, 1.0 & \\
0.13727 & 0.1364 & 570 & 716 & 40 & 0.02, 0.04, 0.06, 0.08, 1.0 &\\
0.13754 & 0.1364 & 411 & 637 & 40 & 0.02, 0.04, 0.06, 0.08, 1.0 &\\
0.13770 & 0.1364 & 296 & 596 & 160 & 0.02, 0.04, 0.06, 0.08, 1.0 &\\
\hline
\hline
\end{tabular}
\caption{Statistics of data set II.}
\label{table:stat_dataset2}
\end{table}

\begin{table}[!ht]
\centering
\begin{tabular}{c c  c  c  c  c  c}
\hline
\hline
&&\multicolumn{5}{c}{$Q^2$ (GeV$^2$), fit range: [11,17]}\\
\hline
$M_\pi$(MeV)  & bin size                 & 0.02      & 0.04      & 0.06       & 0.08       & 0.10        \\
702 & 50$\tau$                          & 0.9818(5) & 0.9645(9) & 0.9473(17) & 0.9308(17) & 0.9155(22)  \\
570 & 50$\tau$                         & 0.9796(6) & 0.9562(17) & 0.9385(23) & 0.9217(27) & 0.9030(31)   \\
411 & 50$\tau$                         & 0.9727(11) & 0.9506(23) & 0.9229(34) & 0.9083(52) & 0.8927(75)  \\
296 & 50$\tau$                         & 0.9733(16) & 0.9462(45) & 0.9221(50) & 0.8911(70) & 0.8959(96)  \\
\hline
\hline
\end{tabular}
\caption{Pion form factor obtained with data set II}
\label{table:PFF_DataSet2}
\end{table}

\section{Pion electromagnetic form factor and charge radius}

\subsection{Monopole analysis of the $Q^2$ dependence of the form factor}

\begin{figure}[!ht]
\centering
\subfigure[$M_\pi \approx 702 \rm MeV$]{\includegraphics[scale=0.5363]{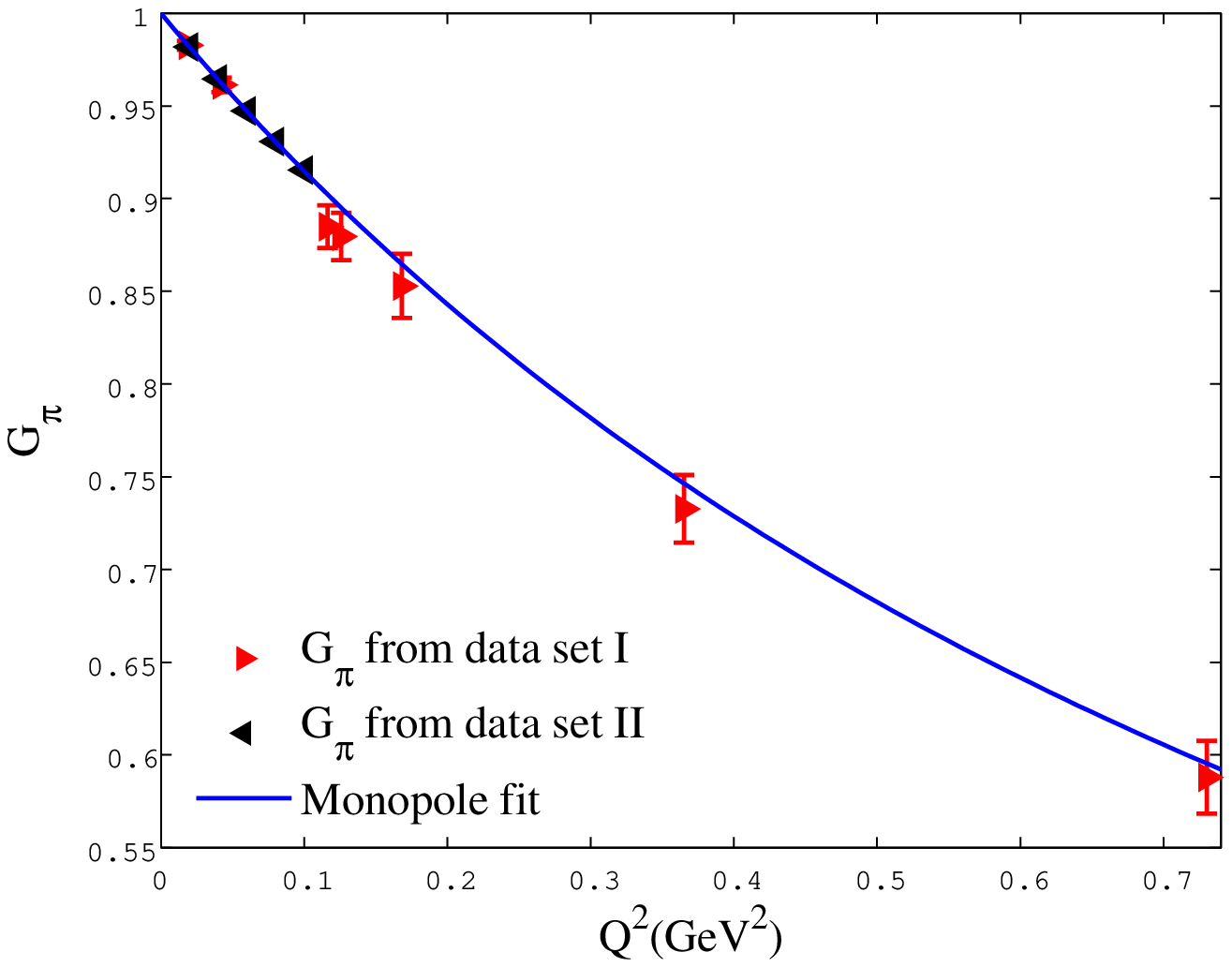}}
\subfigure[$M_\pi \approx 570 \rm MeV$]{\includegraphics[scale=0.5363]{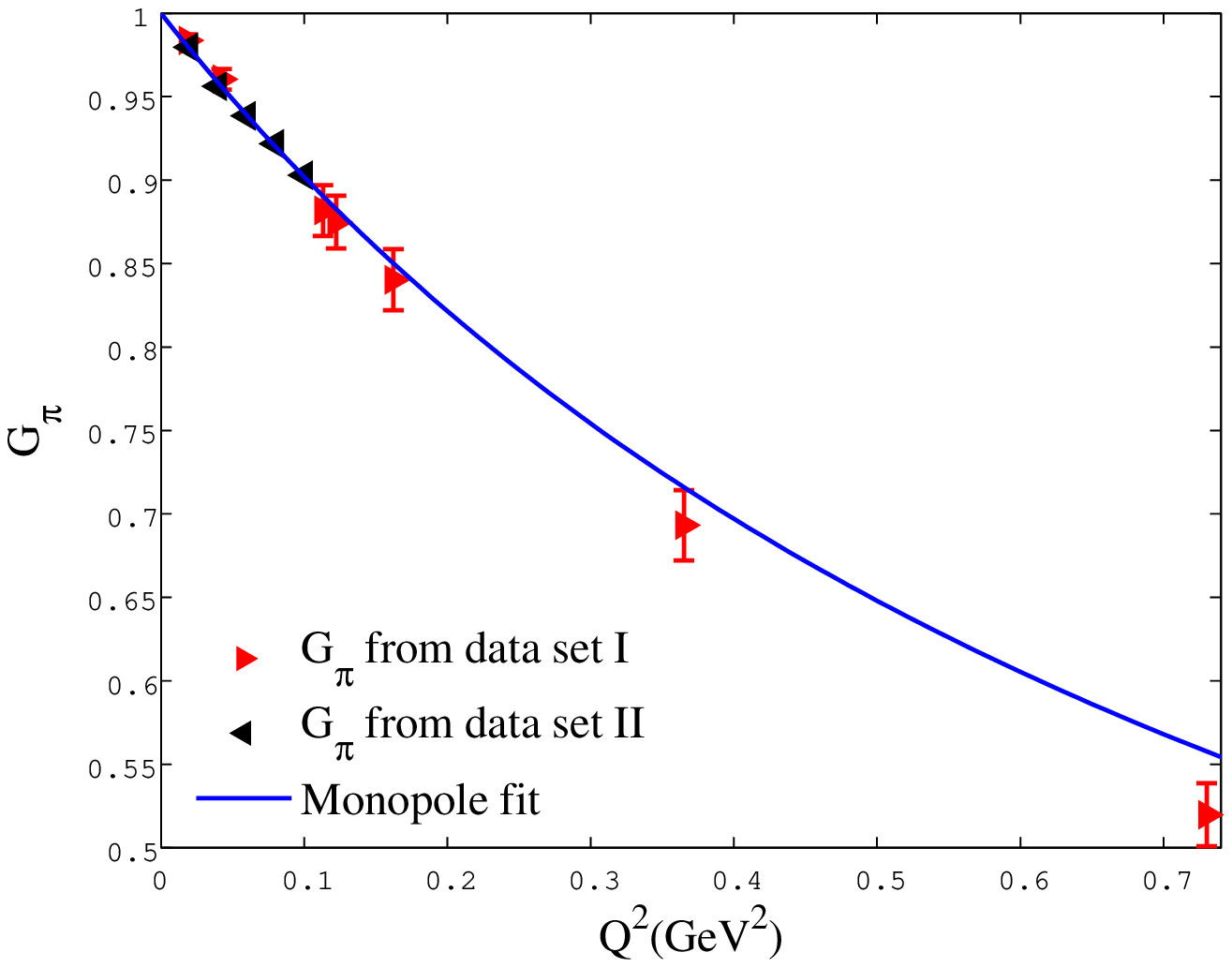}}
\\
\subfigure[$M_\pi \approx 411 \rm MeV$]{\includegraphics[scale=0.5363]{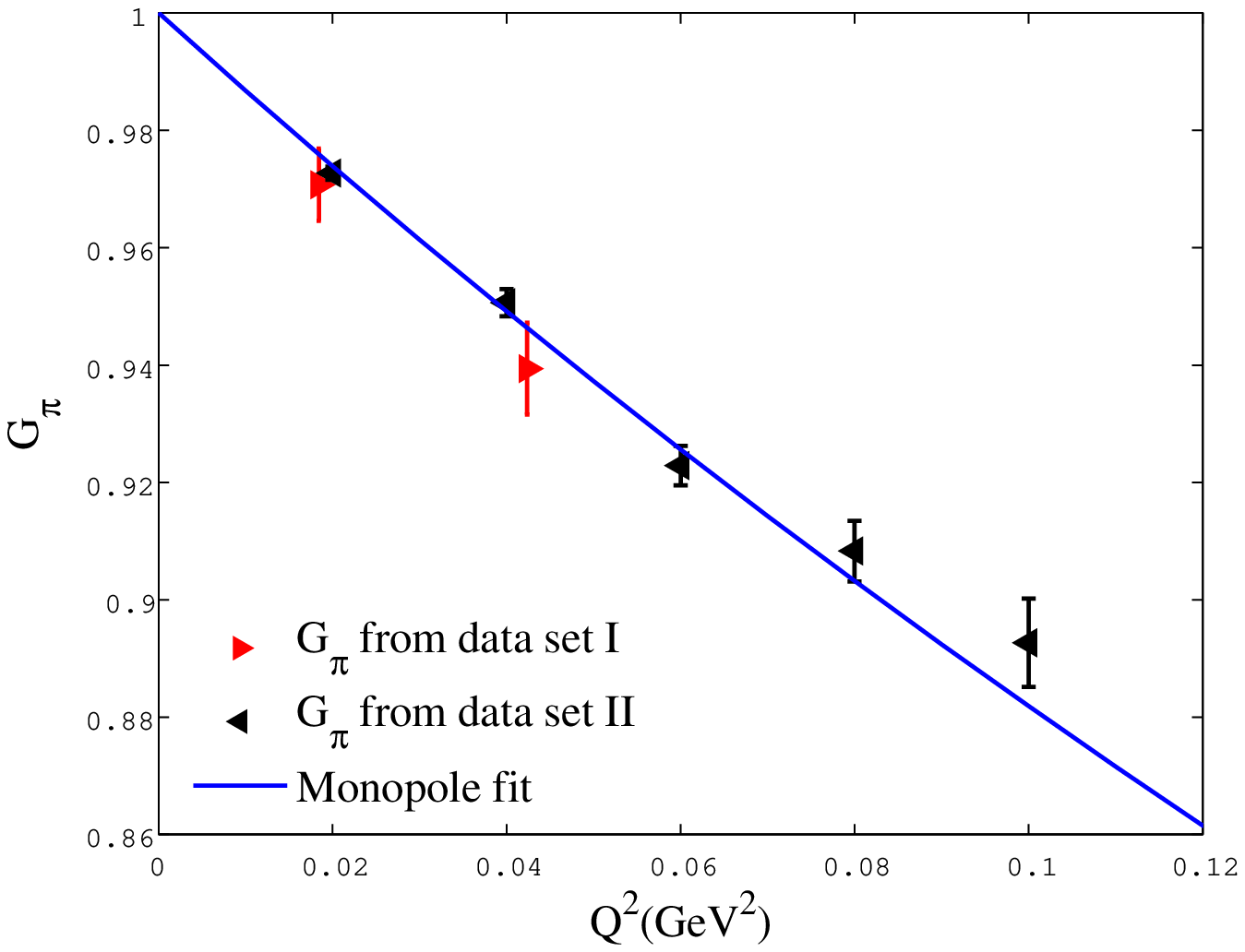}}
\subfigure[$M_\pi \approx 296 \rm MeV$]{\includegraphics[scale=0.5363]{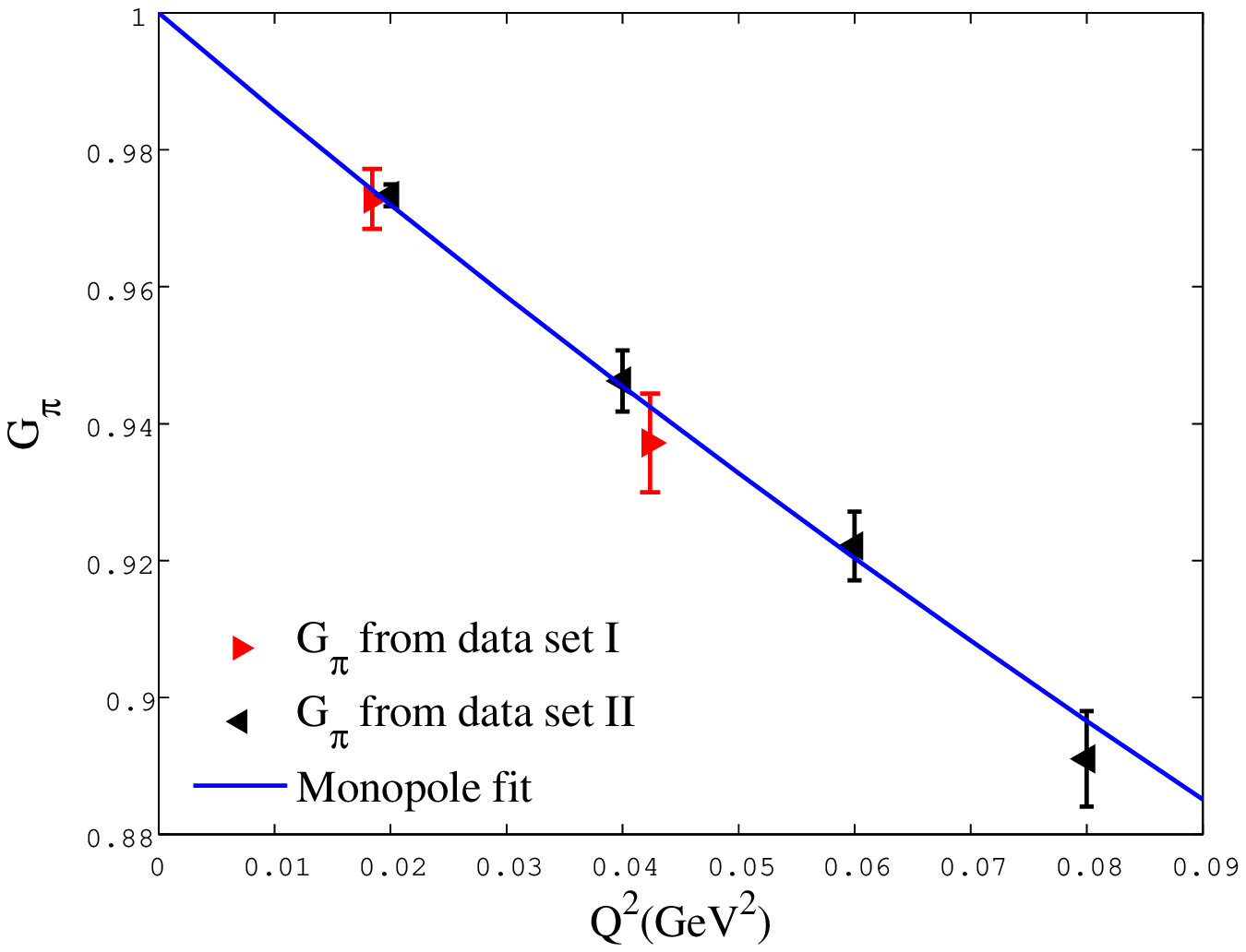}}
\caption{Momentum dependence of $G_\pi(Q^2)$ from data set I and II. 
Blue lines are fits of data I and II to the monopole ansatz (\ref{eq:Mono}).}
\label{fig:PQ2Mono}
\end{figure}

\begin{figure}[!ht]
\centering
{\includegraphics[scale=0.65]{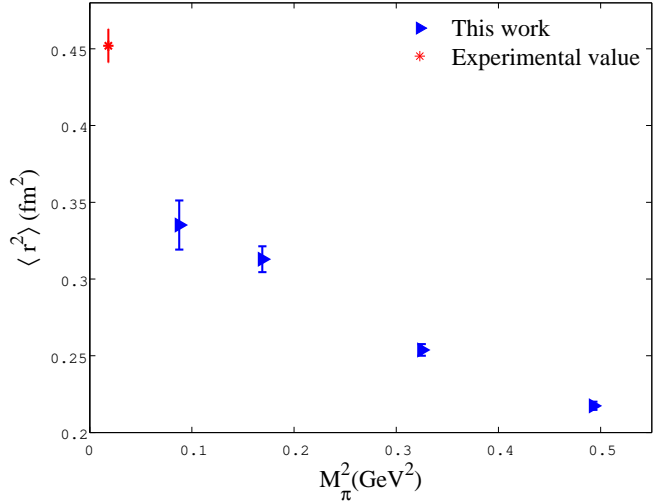}}
\caption{Squared charge radius $\left<r^2\right>(\rm fm^2) $ as a function of $M_\pi^2 $ obtained with the monopole ansatz.  The left-most point 
represents experiment.}
\label{r2_Mpi}
\end{figure}

Figure \ref{fig:PQ2Mono} shows the momentum transfer dependence of our data for the pion form factor at all simulated pion masses. 
The data set I and II are consistent with each other within the estimated errors. 
The experimental pion form factor is phenomenologically reasonably described by a monopole form suggested by the vector meson dominance model,
\begin{equation}
G_\pi(Q^2) = \frac{1}{1+{Q^2/M^2_{mono}}}.
\label{eq:Mono}
\end{equation}
Our data are accordant with the ansatz; solid lines in Fig.~\ref{fig:PQ2Mono} are fits to the monopole form (\ref{eq:Mono}). 
For monopole analysis, we utilize the form factor data in the range up to $Q^2 = 0.08~\rm GeV^2$ at $M_\pi = 296~\rm MeV$ and up to 0.10 $\rm GeV^2$ at 411~MeV, 
since at larger four-momentum transfers plateau signals are not clear. 
The fitted values of the monopole mass $M^2_{mono}$ can be used to estimate the pion electromagnetic charge radius 
{\it via} $\left<r^2\right>=6/M^2_{mono}$.
Results are tabulated in Table \ref{table:BothDataSet_r2} and plotted in Fig.~\ref{r2_Mpi}. The charge radius exhibits an increase as pion mass decreases. 

\begin{table}[!ht]
\centering
\begin{tabular}{c c c c c c c}
\hline
\hline
$\kappa_{ud}$              &  0.13700    & 0.13727    & 0.13754    & 0.13770 \\
$\left<r^2\right>$(fm$^2$) &  0.2174(27) & 0.2538(38) & 0.3129(84) & 0.3352(160)\\
$\chi^2/d.o.f$               &  0.51(12)     & 1.01(12)     & 1.12(19)     & 0.44(11)\\
\hline
\hline
\end{tabular}
\caption{Pion squared charge radius calculated from the monopole fit of data set I and II. 
Errors are estimated by Jackknife method with bin size of 50 $\tau$.}
\label{table:BothDataSet_r2}
\end{table}

\subsection{ChPT analysis to NLO}

For small values of momentum transfer and pseudo-scalar meson masses, we expect ChPT to provide a description 
of the pion form factor as a function of those variables. 
Here we analyze our data in terms of ChPT to NLO.  The analytical expression for the form factor has been worked out long time 
ago both for ${\rm SU(2)_L \times SU(2)_R}$ \cite{gasserleutwyler} 
and ${\rm SU(3)_L \times SU(3)_R}$ \cite{gasserleutwyler_2} cases, which is given by 
\begin{equation}
G_\pi^{SU(2),NLO}(Q^2)  = 1 + \frac{2Q^2}{f^2}l^r_6 + \frac{2M_\pi^2}{f^2} \left[ - \frac{sL}{6} + \frac{H(s)}{N} \right],
\label{PFF_SU2_NLO}
\end{equation}
and 
\begin{equation}
G_\pi^{SU(3),NLO}(Q^2) = 1  - \frac{4Q^2}{f^2_0}L^r_9 + \frac{2M_\pi^2}{f_0^2} \left[  - \frac{sL}{6} + \frac{H(s)}{N} \right] 
                            +  \frac{M_K^2}{f_0^2} \left[ - \frac{s_KL_K}{6} + \frac{H(s_K)}{N} \right],
\label{PFF_SU3_NLO}
\end{equation}
where
\begin{equation}
H(x) = -\frac{4}{3} + \frac{5}{18}x - \frac{x-4}{6} \sqrt{\frac{x-4}{x}}{\rm log} \left( \frac{\sqrt{\frac{x-4}{x}}+1}{\sqrt{\frac{x-4}{x}}-1}\right),
\end{equation}
and $f$ and $f_0$ are the decay constant in the SU(2) and SU(3) chiral limit, respectively. 
In the above equations, we made use of the following definitions:
\begin{eqnarray}
\label{eq:constants}
N &=& (4\pi)^2,\\\nonumber
s &=& \frac{-Q^2}{M_\pi^2}, s_K = \frac{-Q^2}{M_K^2},\nonumber\\
L &=& \frac{1}{N}{\rm log}(\frac{M^2_\pi}{\mu^2}),L_K = \frac{1}{N}{\rm log}(\frac{M^2_K}{\mu^2}).\nonumber
\end{eqnarray}
Besides the decay constant at the chiral limit, 
the SU(2) formula (\ref{PFF_SU2_NLO}) involves $l^r_6$  as the only unknown LEC, and  
the same situation holds for the SU(3) case, (\ref{PFF_SU3_NLO}), with $L^r_9$ as the unknown LEC.
Calculating the slope at the origin of the momentum transfer yields the expressions for the charge radius:
\begin{equation}
\left< r^2\right>_{SU(2),NLO} = -\frac{2}{f^2} \left( 6l^r_6 + \frac{1}{N} + L \right),
\label{r2_SU2_NLO}
\end{equation}
\begin{equation}
\left< r^2\right>_{SU(3),NLO} = -\frac{2}{f_0^2} \left( -12L^r_9 + \frac{3}{2N} + L + \frac{L_K}{2}\right).
\label{r2_SU3_NLO}
\end{equation}

\begin{table}[!t]
\centering
\begin{tabular}{c c c c c c c c c}
\hline
\hline
$M_\pi (\rm MeV)$ & \multicolumn{3}{c}{SU(2)} & \multicolumn{3}{c}{SU(3)}\\
                       & $l^r_6(\mu=1/a)$ & $\chi^2/d.o.f$ & $\left< r^2 \right>_{phys}$(fm$^2$) & $L^r_9(\mu=1/a)$ & $\chi^2/d.o.f$ 
					   & $\left< r^2 \right>_{phys}$(fm$^2$)\\
\hline
 296 & -0.00737(45) & 0.29(10) & 0.366(14) & 0.00256(19) & 0.29(10) & 0.380(14)\\
 411 & -0.00728(26) & 2.34(26) & 0.363(8)  & 0.00260(11) & 2.31(26) & 0.383(8)\\
\hline
\hline
\end{tabular}
\caption{NLO ChPT fit of $G_\pi(Q^2)$ at fixed pion mass $M_\pi$. LEC's are calculated at $\mu=1/a=2.176$~GeV. 
The decay constants in the chiral limit are taken from the work of PACS-CS collaboration \cite{pacscs}:
$f=124.8(5.1) {\rm \ MeV}, f_0=116.0(8.8) {\rm \ MeV}$. Those values are determined with pion masses up to 411 MeV. 
Physical value of the squared charge radius, $\left<r^2\right>_{phys}$, 
is calculated at the physical pion mass for the SU(2) case and at the physical pion and kaon masses for the SU(3) case.}
\label{table:PFF_ChPT_NLO-f}
\end{table}

\begin{figure}[!ht]
\centering
\subfigure[NLO SU(2) ChPT fit of the form factor.]{\includegraphics[scale=0.5363]{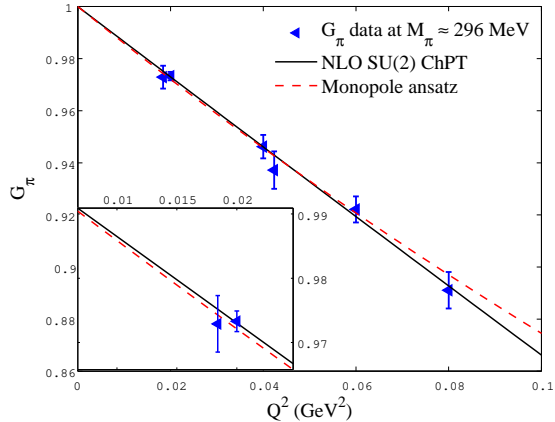}}
\subfigure[NLO SU(3) ChPT fit of the form factor.]{\includegraphics[scale=0.5363]{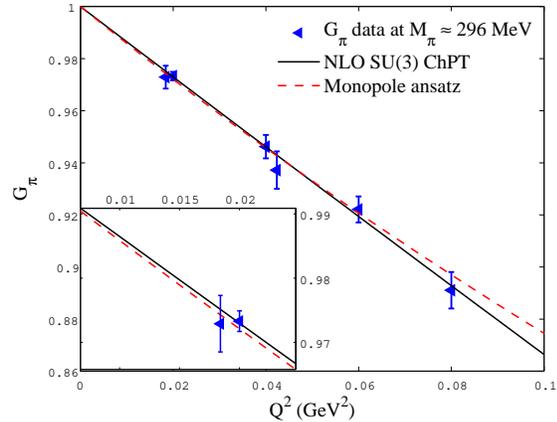}}
\\
\subfigure[$\left<r^2\right>$ determined from the NLO SU(2) ChPT fit of the form factor in comparison with values from the monopole ansatz and the experiment.]
{\includegraphics[scale=0.5363]{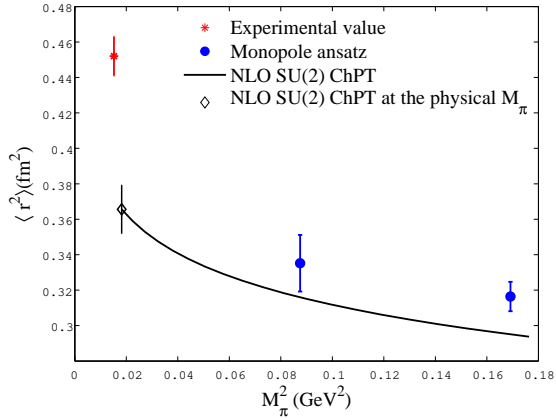}}
\subfigure[$\left<r^2\right>$ determined from the NLO SU(3) ChPT fit of the form factor in comparison with values from the monopole ansatz and the experiment. 
Along the ChPT line displayed in the figure, the kaon mass is fixed at its physical value.]
{\includegraphics[scale=0.5363]{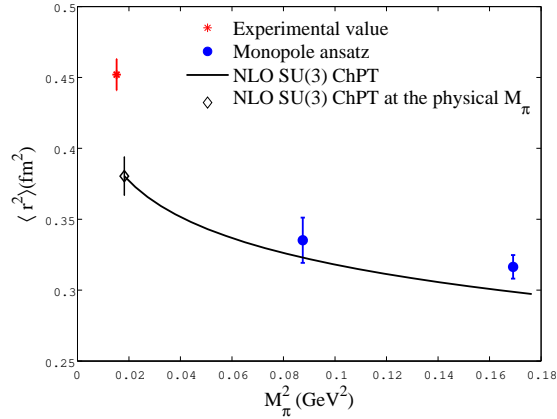}}
\caption{Results of NLO analysis at the pion mass $M_\pi \approx 296 \rm MeV$. 
Fit curves as compared to the form factor data and results for the charge radius are shown for the SU(2) case in panel (a) and (c), 
and for the SU(3) case in panel (b) and (d).} 
\label{fig:R2_ChPT_NLO-f}
\end{figure}

In Table \ref{table:PFF_ChPT_NLO-f} we present results of NLO fits of the form factor for both SU(2) and SU(3) ChPT.  
Fits are made fixing the pion mass at each of the two lightest values, $M_\pi=411$ and $296$~MeV, available in our data set.  
Let us first look at the SU(3) results.  In this case the measured kaon and pion masses are used in the fit, 
while the physical masses are substituted for computing the charge radius at the physical point from the fit results.  
The charge radius extrapolated to the physical point, while consistent within the error for the two pion mass values, 
falls short of the experiment by about 15\%. 
The SU(2) results in Table \ref{table:PFF_ChPT_NLO-f} are similar.  
The value for the charge radius predicted at the physical pion mass is about 20\% smaller than experiment.  
We find similar values in the previous studies \cite{brommel, ETMC, JLQCD, RBC} carried out over a similar range of pion mass. 

In Fig.~\ref{fig:R2_ChPT_NLO-f} we plot the fit curves of the pion form factor 
together with curves from the monopole ansatz for the case of $M_\pi=296$~MeV
for (a) SU(2) and (b) SU(3) ChPT to NLO.  The pion mass dependence of the squared charge radius $\left< r^2\right> (\rm fm^2)$ 
which results from the fits are given in the panels (c) and (d) for the SU(2) and SU(3) cases, respectively.  
Filled circles are the estimates from the monopole ansatz, and the asterisk on the left is the experimental value. 
As indicated in the panels (c) and (d), the NLO ChPT predictions for $\left< r^2\right>$ at 
296~MeV are smaller than that obtained from the monopole ansatz. These differences are also visible in the panels (a) and (b) 
as indicated in the magnified region of small four-momentum transfers.

The NLO ChPT fit at $M_\pi=296$~MeV has a smaller $\chi^2/d.o.f$ compared to that of monopole fit tabulated in 
Table \ref{table:BothDataSet_r2}. 
This is not the case at the pion mass of 411~MeV, however, where $\chi^2/d.o.f$ of the NLO ChPT fit is significantly larger than that of 
the monopole fit.  As we shall discuss below in Fig.~\ref{fig:R2_ChPT_NLO_411_Qsq010}(a) for the SU(2) case,  this is due to 
an upward curvature of the form factor data as $Q^2$ increases to 0.08 and 0.10~GeV$^2$.  
Higher order terms in $Q^2$ need to be included in order to explain the behavior of our data for the form factor at 411~MeV.

\begin{table}[t]
\centering
\begin{tabular}{c c c c c c c c}
\hline
\hline
$M_\pi (\rm MeV)$ & \multicolumn{3}{c}{SU(2)} & \multicolumn{3}{c}{SU(3)}\\
                       & $l^r_6(\mu=1/a)$ & $\chi^2/d.o.f$ & $\left< r^2 \right>_{phys}$(fm$^2$) & $L^r_9(\mu=1/a)$ & $\chi^2/d.o.f$ 
			       	   & $\left< r^2 \right>_{phys}$(fm$^2$)\\
\hline
296 & -0.01238(66) & 0.29(10) & 0.457(18) & 0.00577(33) & 0.29(10) & 0.462(18) \\
411 & -0.01408(45) & 2.38(27) & 0.502(12) & 0.00666(22)  & 2.37(27) & 0.509(12)\\
\hline
\hline
\end{tabular}
\caption{NLO ChPT fit of $G_\pi(Q^2)$ utilizing $M_\pi^2/f_\pi^2$ as the expansion parameter at fixed pion mass $M_\pi$. Values of the decay constant 
at the simulation points are $f_\pi = 151.7(2.7)\rm MeV$ and  162.8(2.6)~MeV at $M_\pi = 296 \rm MeV$ and 411~MeV, respectively.
The physical decay constant, $f^{phys}_\pi = 132.7(5.5)\rm MeV$, is obtained from analysis of data with pion masses up to 411 MeV. 
Physical value of the squared charge radius, $\left<r^2\right>_{phys}$, is calculated at the physical pion mass for the SU(2) case 
and and at the physical pion and kaon masses for the SU(3) case.}
\label{table:PFF_ChPT_NLO-fpi}
\end{table}

\begin{figure}[!ht]
\centering
\subfigure[NLO SU(2) ChPT fit of the form factor employing $M_\pi^2/f^2$ as the expansion parameter.]
{\includegraphics[scale=0.5363]{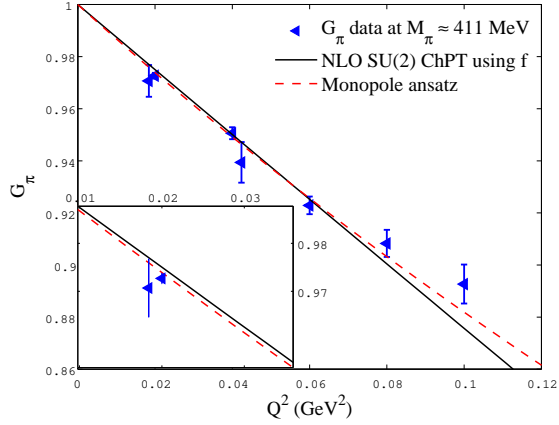}}
\subfigure[NLO SU(2) ChPT fit of the form factor employing $M_\pi^2/f_\pi^2$ as the expansion parameter.]
{\includegraphics[scale=0.5363]{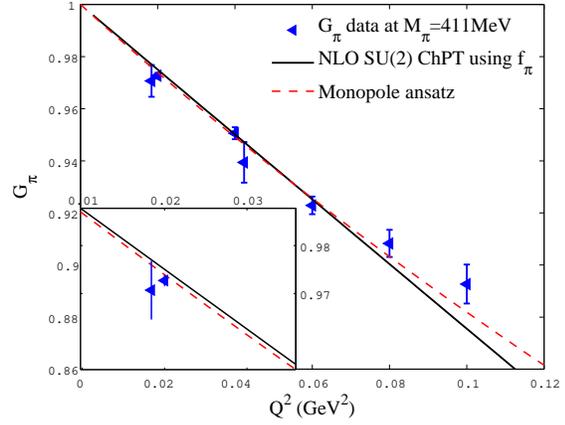}}
\\
\subfigure[$\left<r^2\right>$ determined from the ChPT fit of the form factor utilizing $f$.]
{\includegraphics[scale=0.5363]{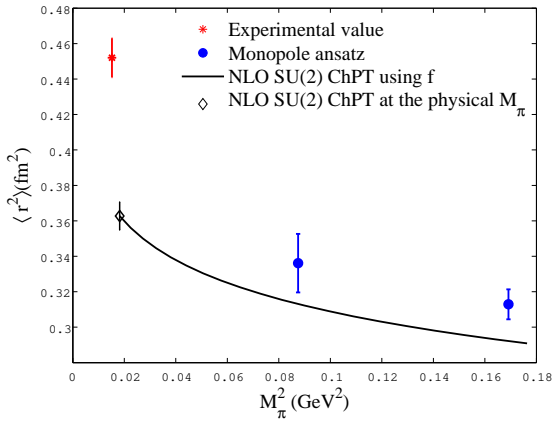}}
\subfigure[$\left<r^2\right>$ determined from the ChPT fit of the form factor utilzing $f_\pi$.]
{\includegraphics[scale=0.5363]{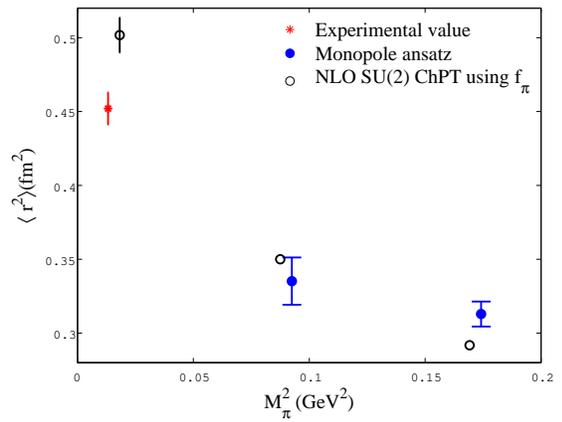}}
\caption{Results of SU(2) NLO analysis at the pion mass $M_\pi \approx 411 \rm MeV$. 
Fit curves as compared to the form factor data and results for the charge radius are shown for the case of using $M_\pi^2/f^2$ as the expansion parameter 
in panel (a) and (c), and for the case of using $M^2_\pi/f_\pi$ in panel (b) and (d).} 
\label{fig:R2_ChPT_NLO_411_Qsq010}
\end{figure}

We now investigate the choice of the decay constant to be used in the NLO ChPT fit.
With an uncertainty of order $O(p^6)$, the decay constant in (\ref{PFF_SU2_NLO}) and (\ref{PFF_SU3_NLO}) can be chosen to be 
$f_\pi$ measured at each pion mass. 
Table \ref{table:PFF_ChPT_NLO-fpi} shows results for NLO fits using $M_\pi^2/f_\pi^2$ as the expansion parameter.
We observe a large difference in the results depending on whether one uses $f$ or $f_\pi$ for the SU(2) case and $f_0$ or $f_\pi$ 
for the SU(3) case. 
The fit results for $l^r_6$ and $L^r_9$ if one uses $f_\pi$ are larger than those of the fit using $f$ and $f_0$ by 40 to 60\%, 
which raises the values of $\left< r^2 \right>_{phys}$ at the physical point by 20 to 30\%.
Predictions for the charge radius from these fits overestimate the experimental value while those employing $f$ underestimate it. 
This uncertainty clearly indicates the importance of $p^6$ terms in the ChPT interpretation of our form factor data at the 
considered range of pion mass.

Comparison of results using $f$ and $f_\pi$ for the SU(2) case at 411~MeV are made in Fig.~\ref{fig:R2_ChPT_NLO_411_Qsq010}. 
The left panels (a) and (c) are results obtained with $f$ while those using $f_\pi$ are shown in the right panels (b) and (d). 
In both cases, NLO ChPT fits do not explain the upward curvature of the form factor data at $Q^2=0.08$ and 0.10~GeV$^2$.

We should note that the SU(2) ChPT analysis requires tuning of the strange quark mass $m_s$ to the physical value, or alternatively, 
the dependence of the SU(2) LEC's on $m_s$ around its physical value has to be determined from data.   
For the Wilson-clover quark action, there is an additional subtlety
that the strange quark mass, as defined {\it via} the PCAC relation, varies with changing up-down quark hopping parameter even if 
the strange quark hopping parameter is kept fixed.   
Our data taken for only one value of the strange quark hopping parameter, however,  is not detailed enough to fully resolve 
the $m_s$ dependence.  We leave such a precise determination of the $m_s$ dependence for future work. 

\subsection{ChPT analysis to NNLO}

The ChPT analysis to NLO presented in the previous subsection indicates that the NLO is not sufficient for the pion mass as large as 
$M_\pi\approx 300-400$~MeV.
Attempts have been made to carry out fits to NNLO of ChPT \cite{ETMC, JLQCD}, and we try this procedure here for the SU(2) case.   

The NNLO formula for the vector form factor is given by \cite{ChPT_NNLO}
\begin{eqnarray}
G_\pi^{SU(2),NNLO}(Q^2) = 1 &+& 2x_2\left[ \frac{1}{6}(s-4)\bar{J}(s)+s\left( -l^r_6 -\frac{L}{6} - \frac{1}{18N} \right) \right] \nonumber\\
&+& 4x_2^2\left( P_V^{(2)}+U_V^{(2)} \right) + O(x_2^3),
\label{eq:Gpi_SU2_NNLO}
\end{eqnarray}
where 
\begin{equation}
x_2=\frac{M_\pi^2}{f_\pi^2},
\label{eq:parameter}
\end{equation}
with $f_\pi$ the decay constant at the pion mass $M_\pi$, which is related to the decay constant $f$ in the SU(2) chiral limit through
\begin{equation}
f_\pi=f\left[1 + 2\frac{M_\pi^2}{f^2}\left(l^r_4-L\right)\right],
\label{eq:decayconstant}
\end{equation}
at NLO of SU(2) ChPT. The two functions $P^{(2)}_V$ and $U^{(2)}_V$ are given by 
\begin{eqnarray}
P^{(2)}_V&=&s\Big[ -\frac{1}{2}k_1+\frac{1}{4}k_2 - \frac{1}{12}k_4 + \frac{1}{2}k_6 - l^r_4 \left( 2l^r_6 + \frac{1}{9N} \right)
+\frac{23}{36}\frac{L}{N} + \frac{5}{576N} + \frac{37}{864N^2} + r^r_{V1}\Big]\nonumber\\
&&+ s^2\Big[\frac{1}{12}k_1 - \frac{1}{24}k_2 + \frac{1}{24}k_6
+\frac{1}{9N} \left( l^r_1-\frac{1}{2}l^r_2 + \frac{1}{2}l^r_6 - \frac{1}{12}L - \frac{1}{384} - \frac{47}{192N}\right) + r^r_{V2}\Big]
\end{eqnarray}
\begin{eqnarray}
U^{(2)}_V&=&\bar{J} \Big[
 \frac{1}{3}l^r_1(-s^2+4s) + \frac{1}{6}l^r_2(s^2-4s) + \frac{1}{3}l^r_4(s-4) + \frac{1}{6}l^r_6(-s^2+4s)\\\nonumber
 &&-\frac{1}{36}L(s^2+8s-48)+\frac{1}{N}\left( \frac{7}{108}s^2 - \frac{97}{108}s + \frac{3}{4} \right)\Big]\\\nonumber
&&+\frac{1}{9}K_1(s) + \frac{1}{9}K_2(s)\left( \frac{1}{8}s^2-s+4 \right) + \frac{1}{6}K_3(s) \left( s-\frac{1}{3} \right) -\frac{5}{3}K_4(s)
,
\end{eqnarray}
and the integral functions $\bar{J},K_1,K_2,K_3,K_4$ are defined by 
\begin{equation}
\left[
\begin{array}{c}
\bar{J} \\
K_1\\
K_2\\
K_3\\
\end{array}
\right]
=\left[
\begin{array}{cccc}
0        & 0   & z              & -4N   \\
0        & z   & 0              & 0     \\
0        & z^2 & 0              & 8     \\
Nzs^{-1} & 0   & \pi^2(Ns)^{-1} & \pi^2 \\
\end{array}
\right]
\left[
\begin{array}{c}
h^3 \\
h^2 \\
h \\
-(2N^2)^{-1} \\
\end{array}
\right],
\end{equation}
and
\begin{equation}
K_4=\frac{1}{sz}\left( \frac{1}{2}K_1 + \frac{1}{3}K_3 + \frac{1}{N}\bar{J} + \frac{(\pi^2-6)s}{12N^2} \right)
,
\end{equation}
where
\begin{equation}
h(s)=\frac{1}{N\sqrt{z}} ln\frac{\sqrt{z}-1}{\sqrt{z}+1}, \qquad z = 1-\frac{4}{s}.
\end{equation}
As well as notations in (\ref{eq:constants}), we also use
\begin{eqnarray}
\label{eq:constants2}
k_i &=& [4l^r_i-\gamma_iL] L,
\end{eqnarray}
where
\begin{eqnarray}
\label{eq:gamma_consts}
\gamma_1=1/3, \gamma_2=2/3, \gamma_4=2, \gamma_6=-1/3.
\end{eqnarray}
From (\ref{eq:Gpi_SU2_NNLO}) the NNLO ChPT expression for the squared charge radius reads,
\begin{eqnarray}
\left< r^2 \right>_{SU(2),NNLO} &=& -\frac{2}{f_\pi^2}\left( 6l^r_6 + L + \frac{1}{N} \right)\\\nonumber
&+& \frac{4M_\pi^2}{f_\pi^4} \left[ -3k_1 + \frac{3}{2}k_2 - \frac{k_4}{2} + 3k_6 -12l^r_4l^r_6 
+ \frac{1}{N} \left( -2l^r_4 + \frac{31}{6}L + \frac{13}{192} - \frac{181}{48N} \right) + 6r^r_{V1}
 \right].
\end{eqnarray}

\begin{table}[t]
\centering
\begin{tabular}{c c c c c c c c c}
\hline
\hline
exp. parameter & $M_\pi(\rm MeV)$ & $l^r_6(\mu=1/a)$ & $10^4 r^r_{V1}$ & $10^4 r^r_{V2}$ & $\chi^2/d.o.f$ & $\left< r^2 \right>_{phys}$\\
\hline
$M_\pi^2/f^2$         & 296, 411     & -0.0098(11)   & 1.67(51)   & 1.04(60)   & 0.63(9) & 0.420(31) \\ 
\hline
$M_\pi^2/f_\pi^2$     & 296, 411     & -0.0103(18)  & 3.4(1.8) & 3.3(1.6) & 0.72(10) & 0.441(44)\\ 
\hline
\hline
\end{tabular}
\caption{NNLO ChPT SU(2) fit of $G_\pi(Q^2)$ using data at 2 lightest pion masses. The result in the first row is obtained by using 
$f=124.8(5.1)$~MeV in the SU(2) chiral limit \cite{pacscs}, while that in the second row by substituting the measured values of $f_\pi$. }
\label{table:PFF_ChPT_SU2_NNLO_2lightest}
\end{table}

For checking the convergence at NNLO, we carry out fits employing both $M_\pi^2/f^2$ and $M_\pi^2/f_\pi^2$ as the expansion parameter.
For the former fit we use (\ref{eq:decayconstant}) to reexpand the expression for the form factor to the necessary order, 
and use the value of $f$ obtained in \cite{pacscs}.  
Besides the pion decay constant the ChPT formula of the form factor to NNLO depends on 5 other LECs: 
$l^r_1-l^r_2/2,l^r_4,l^r_6,r^r_{V1},r^r_{V2}$. It is very difficult to find a stable fit in the 5-dimension parameter space.
Therefore we fix $l^r_1,l^r_2$ at the phenomenology values since they were calculated with small error bar 
from experimental data\cite{ChPT_NNLO_4}. For $l^r_4$, which is only required in the formulation with $f_\pi$, 
we apply the value obtained by an NLO fit of data in the range $M_\pi\leq 411$~MeV 
by the PACS-CS collaboration\cite{pacscs}. 

\begin{figure}[!ht]
\centering
\subfigure[NNLO SU(2) ChPT fit of the form factor utilizing $f$.]
{\includegraphics[scale=0.5363]{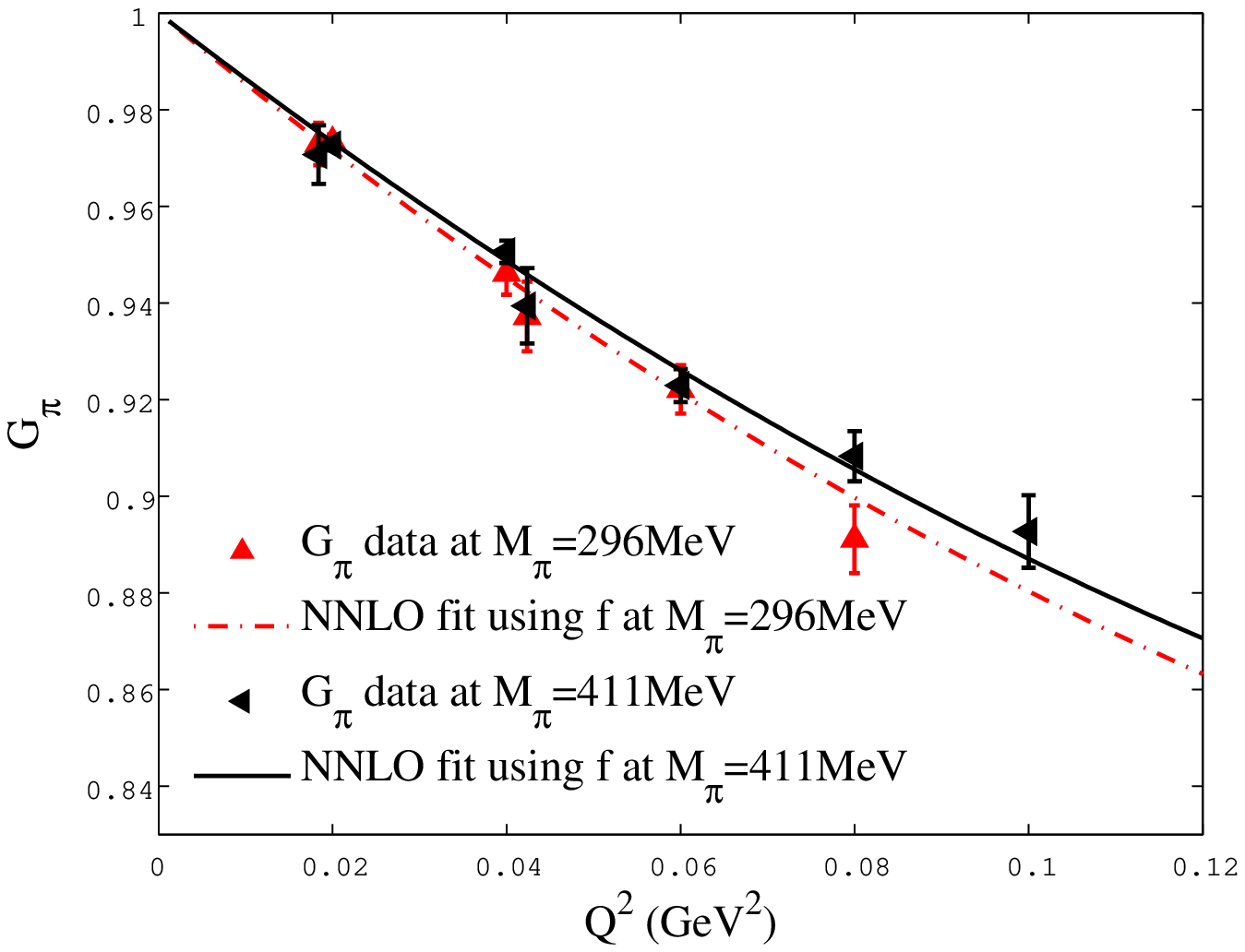}}
\subfigure[$\left<r^2\right>$(fm$^2$) calculated from the fit employing $f$.]
{\includegraphics[scale=0.5363]{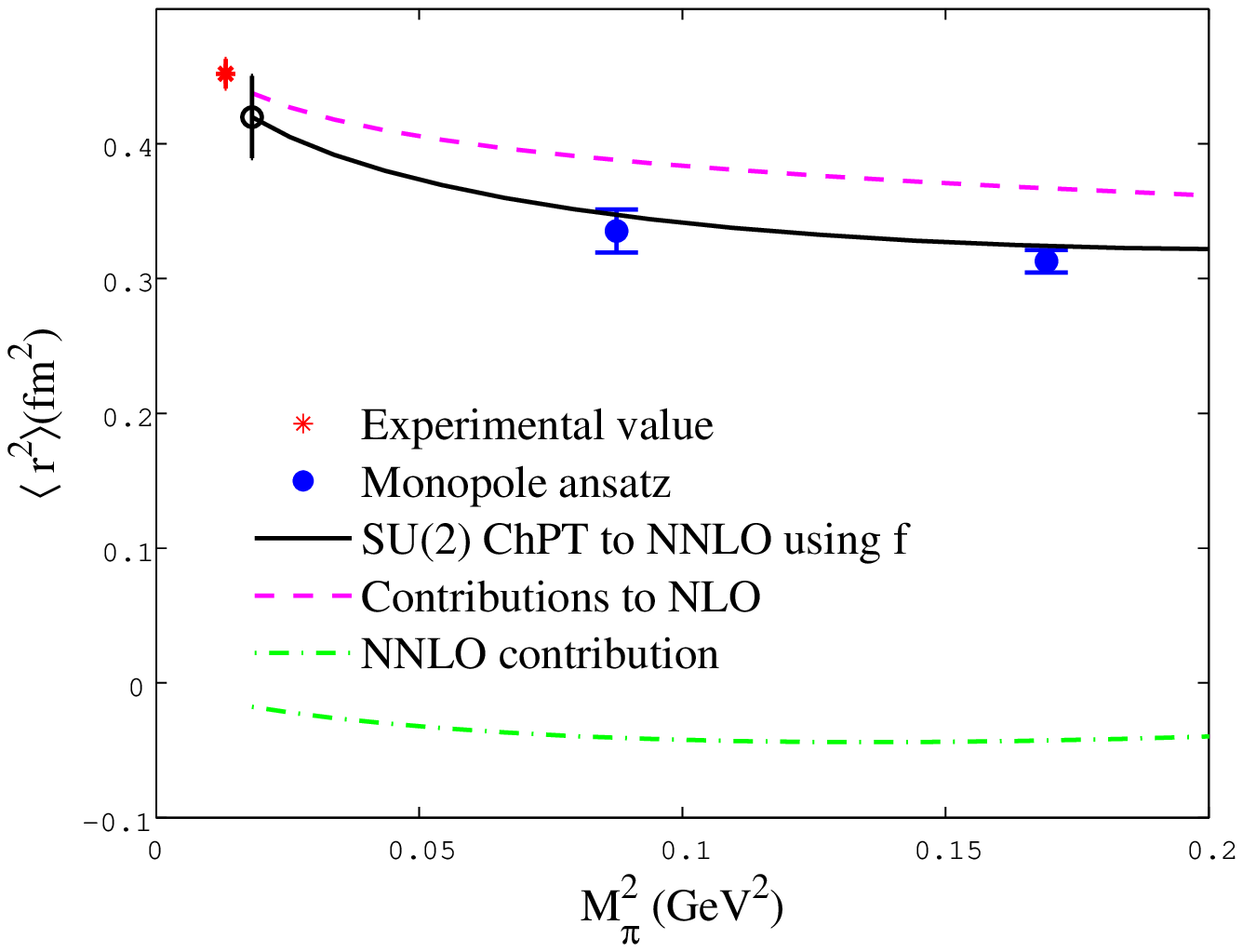}}
\caption{Results of NNLO SU(2) ChPT analysis using the decay constant $f$ in the SU(2) chiral limit and combining data at $M_\pi=296$~MeV and 411~MeV.
In panel (b) filled circles are values estimated from monopole fits, and open circles and lines are fit results. 
For the latter NLO and NNLO contributions are 
also plotted.}
\label{fig:ChPT_SU2_NNLO_2lightest-f}
\end{figure}

\begin{figure}[!ht]
\centering
\subfigure[NNLO SU(2) ChPT fit of the form factor utilizing $f_\pi$.]{\includegraphics[scale=0.5363]{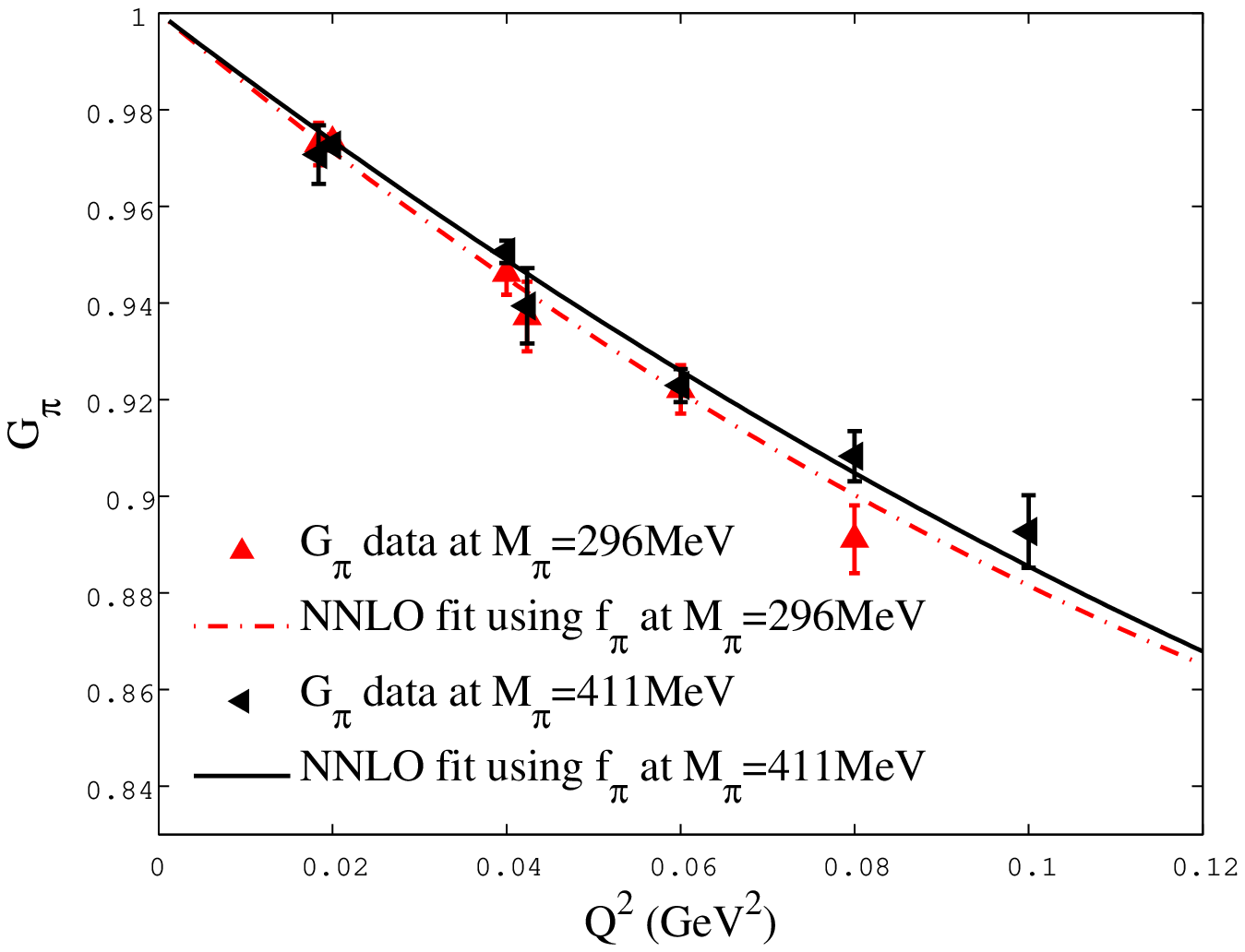}}
\subfigure[$\left<r^2\right>$(fm$^2$) calculated from the fit employing $f_\pi$.]
{\includegraphics[scale=0.5363]{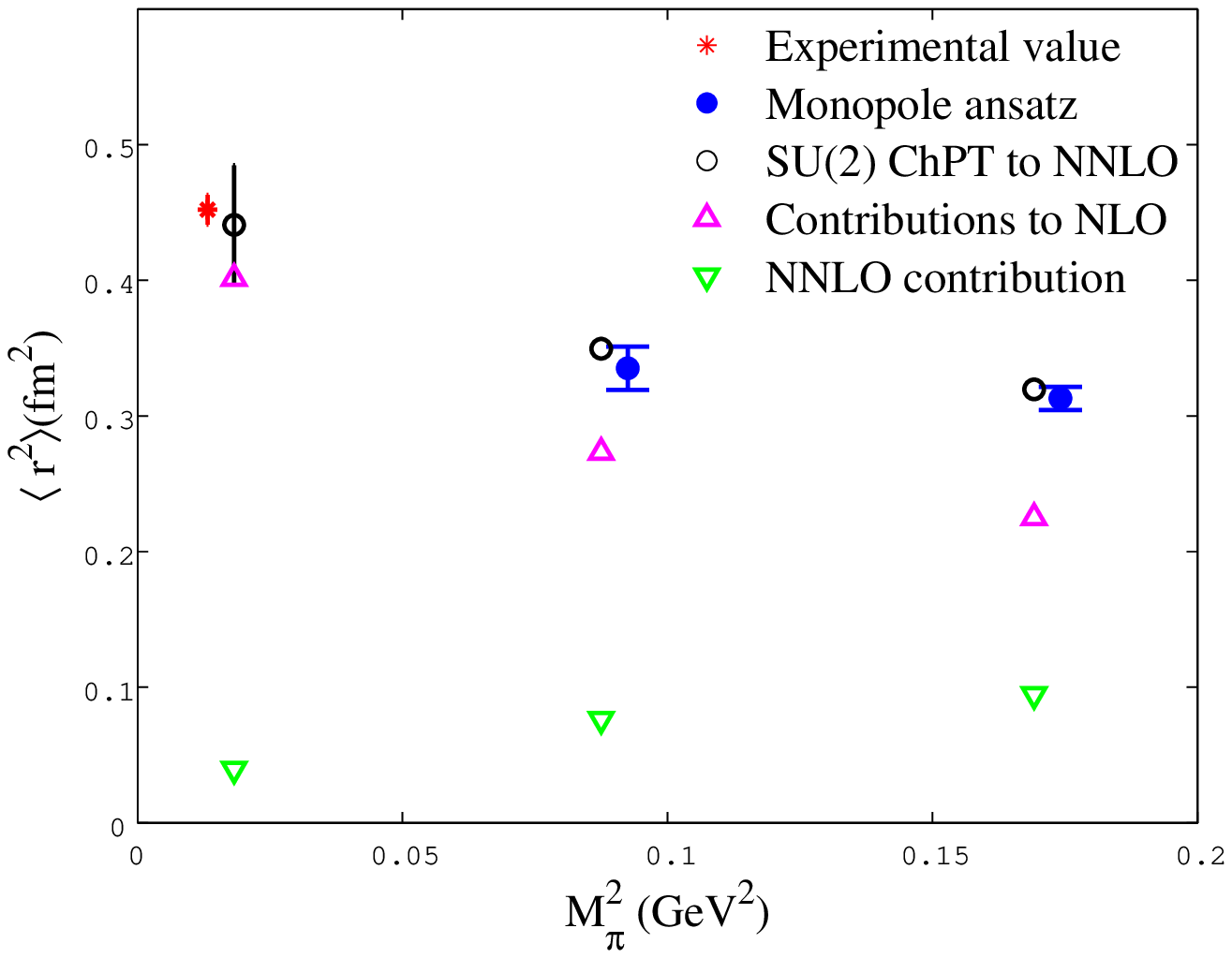}}
\caption{Results of NNLO SU(2) ChPT analysis using the decay constant $f_\pi$ measured at each $M_\pi$ and combining data at $M_\pi=296$~MeV and 411~MeV.
In panel (b) filled circles are estimations from monopole fits, and open circles are fit results. Upward triangles and downward triangles are contributions to NLO 
and NNLO contribution, respectively.}
\label{fig:ChPT_SU2_NNLO_2lightest-fpi}
\end{figure}

We find that stable fits are difficult to obtain unless we utilize data at more than a single pion mass in the fit procedure. 
The fit result obtained with data for the two pion masses $M_\pi=296$ and 411~MeV, which were used for the NLO analysis, is listed in 
Table \ref{table:PFF_ChPT_SU2_NNLO_2lightest} and shown in Figs.~\ref{fig:ChPT_SU2_NNLO_2lightest-f} and \ref{fig:ChPT_SU2_NNLO_2lightest-fpi}. 
Combining the data at the two pion masses is acceptable since strange quark mass does not vary much,  $m_s=89.8(1.3)$ and 92.2(1.3)~MeV at $M_\pi=296$ 
and 411~MeV, respectively \cite{pacscs}.  
We observe that the results for $l^r_6$ are consistent between the two fits within the error of 10--15\% and so are the predictions 
for the squared charge radius at the physical point, indicating that ChPT reasonably converges at NNLO up to $M_\pi\approx 400$~MeV and 
$Q^2\approx 0.01$~GeV$^2$.  This is also seen by plotting the NLO and NNLO contributions separately as shown in 
Figs.~\ref{fig:ChPT_SU2_NNLO_2lightest-f} and \ref{fig:ChPT_SU2_NNLO_2lightest-fpi}.  
The squared charge radius predicted at the physical point is close to the experimental value and is consistent within statistics errors of 10\%. 

\section{Toward the physical point -- a $64^4$ lattice calculation -- }

While ChPT to NNLO yields a reasonable result for the physical pion charge radius, the estimated error of 10\% is quite large. 
We feel that for a convincing understanding of the physical pion charge radius one needs to explore the region of 
pion mass closer to the physical point than the value $M_\pi\approx 300$~MeV analyzed so far.

The PACS-CS gauge configurations has one more set corresponding to $M_\pi\approx 156\rm \ MeV$.  
We tried to calculate the form factor on this set, and found that the pion two- and three-point correlators exhibit very large fluctuations, 
to the extent that taking a meaningful statistical average is difficult. 
This trend becomes more pronounced as the twist carried by quarks becomes larger.  
Since $LM_\pi\approx 2.3$ at this pion mass for $L=32$, we suspect that this phenomenon is caused by a small size of the lattice 
relative to the pion mass, and consequent increase of large fluctuations.  

A natural remedy to this difficulty is to employ larger lattices as one moves toward the physical pion mass. 
PACS-CS collaboration has been pushing a simulation on the physical point on a $64^4$ lattice as a continuation of the work 
on a $32^3\times 64$ lattice.  The hopping parameter of the run is adjusted to the best  estimate of the physical point
$(\kappa_{ud}, \kappa_s) = ( 0.137785, 0.13665)$. We have used a subset of those configurations to calculate the pion form factor 
on a $64^4$ lattice.  This requires much computer time, and hence we have only 4 configurations measured so far. 
We used the same setup as for the data set II, namely, (i) the incoming and outgoing pions in the three-point function 
carry momenta of the same magnitude $|\vec{p'}|=|\vec{p}|$ but point in different directions, 
(ii) 4 random $Z(2) \otimes Z(2)$ wall sources located at $t=0,16,32,48$ are employed, 
(iii) the twist technique is applied to the two quarks running from the source to the current and from the current to the sink, 
and (iv) four values are chosen for the twist angle $\vec{\theta}=\left(\theta, 0, 0\right)$ 
and its permutations such that the four-momentum transfer of the current takes the value $Q^2({\rm GeV}^2)=0.02, 0.04, 0.06, 0.08$. 
The fixed sink time $t_f$ is chosen to be 28.

\begin{figure}[!ht]
\centering
\includegraphics[scale=0.5363]{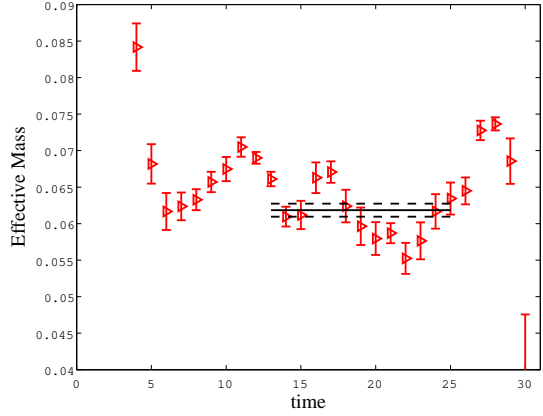}
\caption{Triangle symbols represent effective mass obtained from measurement of 4 configurations of a size $64^4$ with 4 source points located at $t=0,16,32,48$ 
and 4 random sources for each source location. Center lines exhibit PACS-CS estimation of pion mass calculated from larger statistics of 53 configurations.}
\label{fig:C2ptData_64latt}
\end{figure}

Since we use 4 random sources for each of the 4 locations of the source in time, our measurement on 
4 configurations gives 64 two-point functions.  
The pion effective mass from our measurement together with the PACS-CS estimate of pion mass calculated from larger statistics 
of 53 configurations is shown in Fig.~\ref{fig:C2ptData_64latt}.  Although only 4 configurations have been used, 
one can already observe a plateau-like behavior for pion effective mass in this figure. The central value from the PACS-CS estimate 
corresponds to $M_\pi\approx 135$~MeV.  This is somewhat smaller than the charged pion mass, and significantly smaller than 
$M_\pi=156$~MeV considered earlier on a $32^3\times 64$ lattice where we encountered problem of large fluctuations.  

\begin{figure}[!ht]
\centering
\subfigure[Ratio $R'(\tau)$ defined in (\ref{eq:ratio3ptOnly}) at $Q^2(\rm GeV^2)=0.02, 0.04$.]{\includegraphics[scale=0.5363]{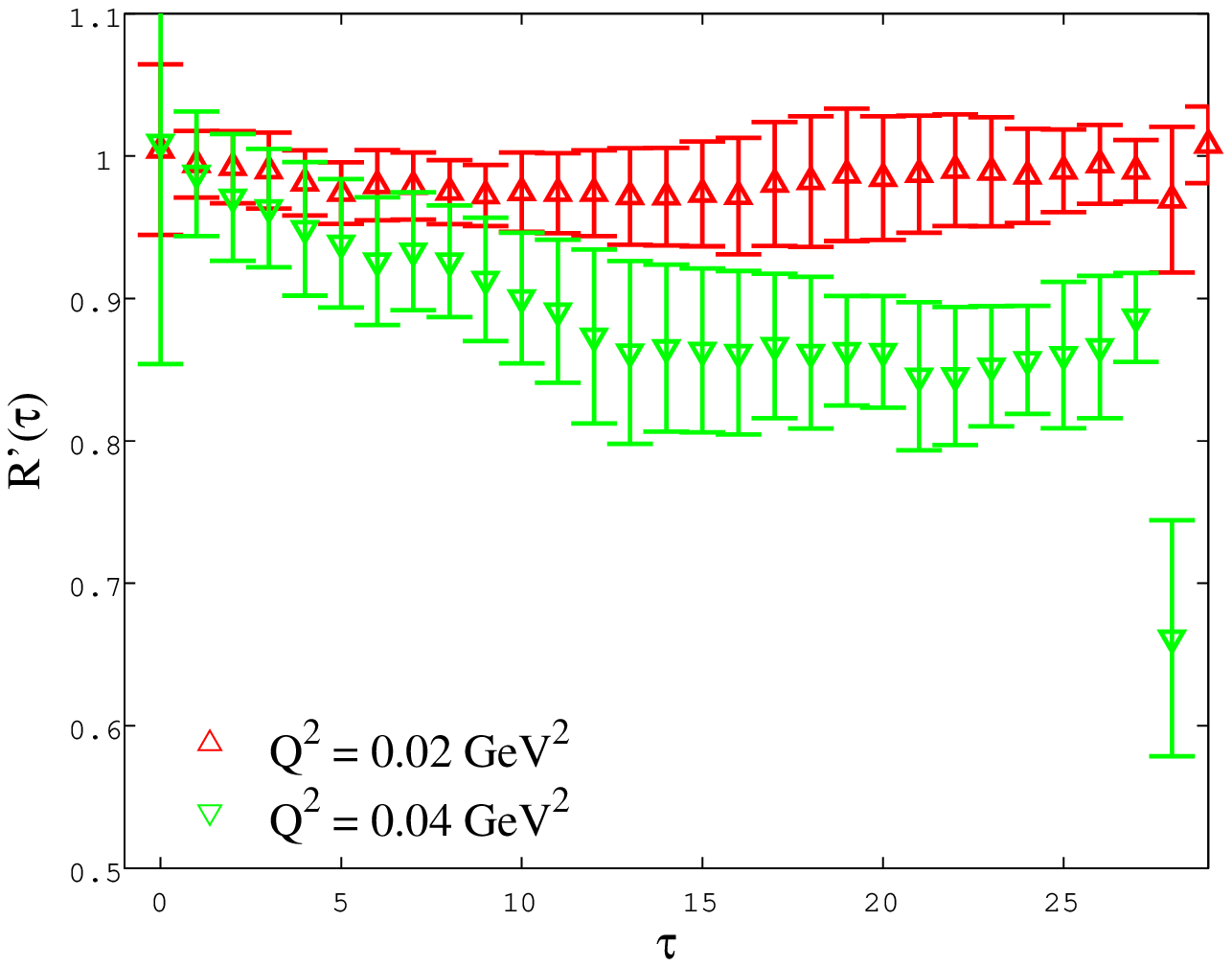}}
\subfigure[Ratio $R'(\tau)$ defined in (\ref{eq:ratio3ptOnly}) at $Q^2=0.06$~GeV$^2$.]{\includegraphics[scale=0.5363]{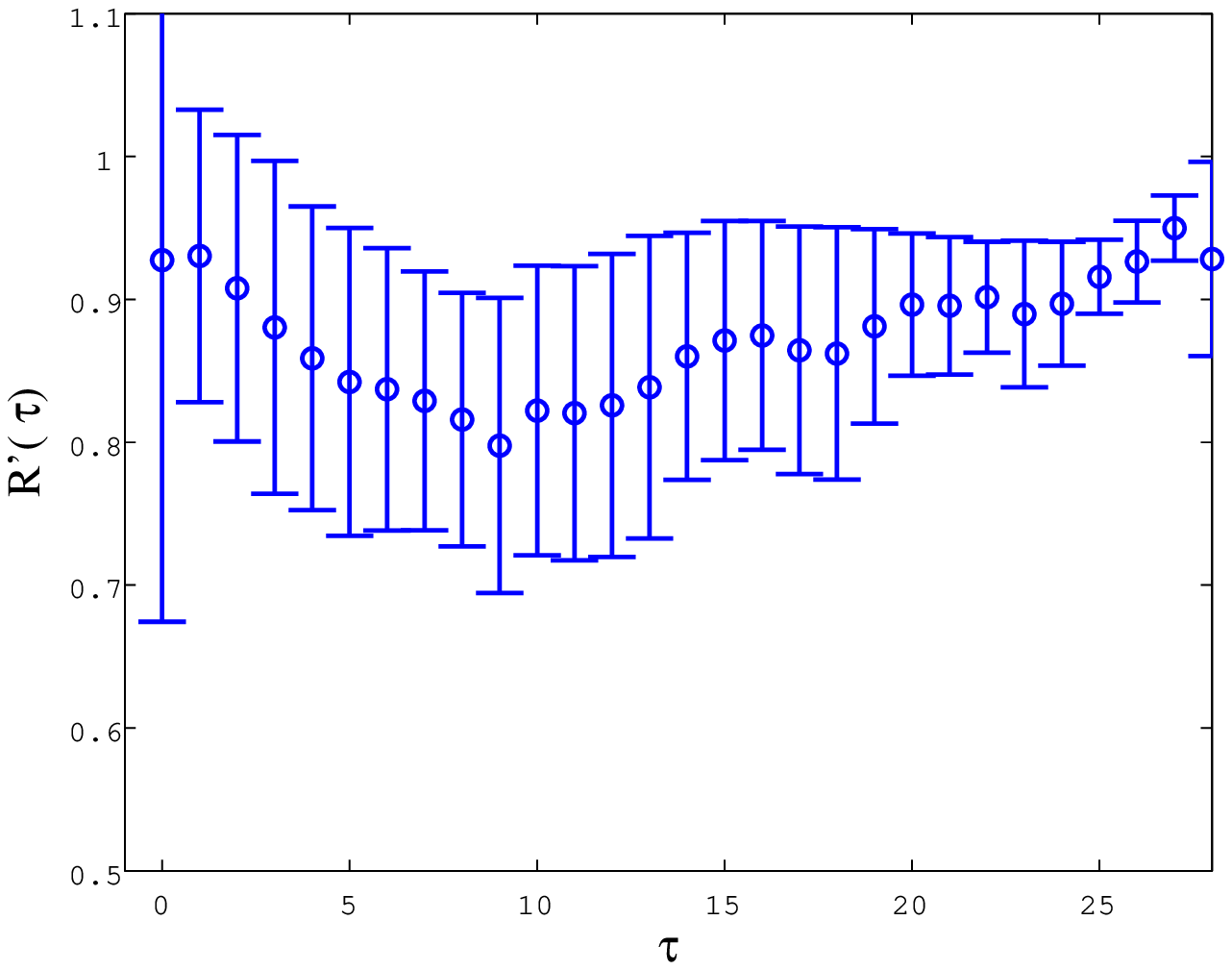}}
\caption{Ratio for extracting the form factor at several $Q^2$ values measured on 4 configurations of a $64^4$ lattice. 
Fitting range is chosen to be $\tau=12 - 16$.}
\label{fig:PFF_64latt}
\end{figure}
\begin{figure}[!ht]
\centering
\includegraphics[scale=0.5363]{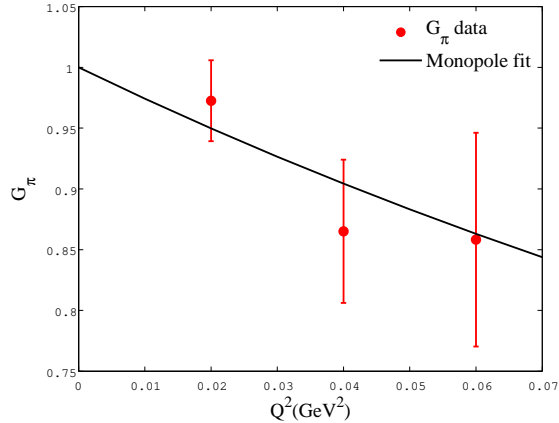}
\caption{Momentum dependence of the form factor at $M_\pi\approx 135$~MeV with measurement taken on 4 configurations of a $64^4$ lattice.
Black line is a fit of two data points closest to $Q^2=0$ to the monopole ansatz (\ref{eq:Mono}).}
\label{fig:PFF_Q2_64latt}
\end{figure}

In Fig.~\ref{fig:PFF_64latt} we plot the ratio $R'(\tau)$ obtained from the 4 configurations. 
Making a constant fit over $\tau=12-16$ yields the result for the form factor displayed in Fig.~\ref{fig:PFF_Q2_64latt}. 
Our data seems reasonable up to $Q^2=0.06~$GeV$^2$.  Estimating the slope at the origin by a monopole fit 
to the two points closest to $Q^2=0$, we obtain $\left< r^2 \right>=0.675(285) {\rm fm}^2$.  
While the error is too large to seriously discuss consistency with experiment, it is certainly encouraging that 
the value is larger than those obtained at $M_\pi$ of about 300~MeV, and 
that the physical point simulation appears possible for the pion electromagnetic form factor on a $64^4$ lattice. 

\section{Conclusion}

We have presented a lattice calculation of the pion electromagnetic form factor in 2+1 dynamical flavor QCD 
with the O(a)-improved Wilson-clover quark action and Iwasaki gauge action.

In order to obtain data with reasonable error at light quark masses close to the physical point, 
we have utilized some improved techniques besides traditional methods for the form factor calculation. 
We have shown that, choosing momenta of the incoming and outgoing pions to have the same magnitude but different directions, 
the ratio for extracting the pion form factor becomes statistically much better behaved. 
We have confirmed the validity of the twisted boundary condition and employed it to explore the form factor in 
the region of small four-momentum transfer. 
Application of the random $Z(2) \otimes Z(2)$ wall source has helped us to save computing time considerably.

\begin{figure}[htb]
\begin{center}
\includegraphics[scale=0.65]{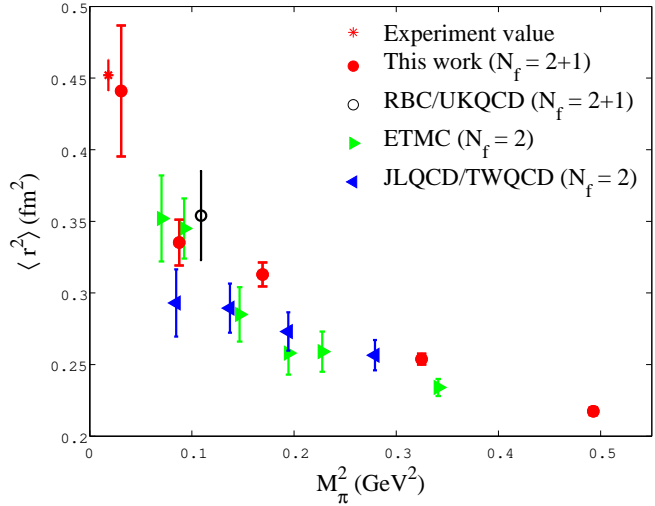}
\end{center}
\caption{$\left< r^2 \right>(\rm fm^2)$ in comparison with previous studies. 
The left-most filled circle represents our NNLO SU(2) ChPT prediction at the physical pion mass.}
\label{fig:6464_r2}
\end{figure}

Our data for the pion mean-square charge radius $\left< r^2 \right>$ agree with recent data of other groups, and show that 
$\left< r^2 \right>$ increases toward the physical value as $M_\pi$ decreases. 
Nevertheless, on a $32^3\times64$ lattice, we could extract reasonable data only down to $M_\pi \approx 296$~MeV. 

ChPT analysis of the form factor utilizing NLO SU(3) or SU(2) formula lead to the squared charge radius which is smaller than experiment  
by 15--20\%.  Employing ChPT to NNLO improves the agreement. In fact our NNLO SU(2) fit using $M_\pi^2/f_\pi^2$ as the expansion parameter yields 
$\left<r^2\right>_{phys} = 0.441(44)(13)(\rm fm^2)$ where the first error is statistical and the second error due to the error in the lattice spacing.

We feel that a complete explanation of the behavior of the squared charge radius would require successful calculation of the form factor below 
$M_\pi\approx 300$~MeV. Our experience points toward the necessity of enlarging the lattice size sufficiently.  A sample calculation on a $64^4$ 
lattice indicates that the $64^4$ lattice with $M_\pi L\approx 4$ probably satisfies the requirement.  
We leave further exploration of the form factor calculation on such a lattice as future work. 

\begin{acknowledgments}
Numerical calculations for the present work have been carried out
under the ``Interdisciplinary Computational Science Program'' of 
Center for Computational Sciences, University of Tsukuba. 
We thank the members of the PACS-CS Collaboration for discussions. 
This work is supported in part by Grants-in-Aid for Scientific Research
from the Ministry of Education, Culture, Sports, Science and Technology
(Nos.
16740147,   %Ishikawa
18104005,   %Ukawa
20740123,   %Ukita
20740139    %Ishikawa
).
\end{acknowledgments}


\clearpage
\begin{thebibliography}{99}

%pion form factor experiment

%\bibitem{na7} 
%NA7 Collaboration, S. R. Amendolia {\it et al.}, \emph{A measurement of space-like pion electromagnetic form factor, Phys. Lett.} {\bf B146}, 116 (1984).

\bibitem{pdg} 
K. Nakamura {\it et al.} (Particle Data Group), \emph{Review of Particle Physics, J. Phys. } {\bf G37}, 075021 (2010).

\bibitem{gasserleutwyler}
J. Gasser and H. Leutwyler, \emph{Chiral perturbation theory to one loop}, \emph{Ann. Phys.} {\bf 158}, 142 (1984)

\bibitem{gasserleutwyler_2}
J. Gasser and H. Leutwyler, \emph{Chiral perturbation theory: expansions in the mass of the strange quark}, \emph{Nucl. Phys.} {\bf B250} (1985) 465.


\bibitem{martinelli}
G. Martinelli, Christopher T. Sachrajda,, \emph{A Lattice Calculation of the Pion's Form-Factor and Structure Function}, \emph{Nucl. Phys.}, 
{\bf B306}, 865 (1988)

\bibitem{draper}
T. Draper, R. M. Woloshyn, W. Wilcox, and K.-F. Liu, \emph{The pion form factor in lattice QCD , Nucl. Phys.} {\bf B318}, 319 (1989)

%%%%%%%%%%%%%%%%%%%%%%%%%%%

\bibitem{brommel}
D. Brommel et al. (QCDSF/UKQCD Collaboration), \emph{The Pion form-factor from lattice QCD with two dynamical flavours}, \emph{Eur. Phys. J.}, 
{\bf C51}, 335 (2007)

\bibitem{ETMC}
R. Frezzotti et al., \emph{Electromagnetic form factor of the pion from twisted-mass lattice QCD at N(f) = 2}, \emph{Phys. Rev. D} {\bf 79}, 
074506 (2009)

\bibitem{JLQCD}
S. Aoki et al. (JLQCD/TWQCD Collaboration), \emph{Pion form factors from two-flavor lattice QCD with exact chiral symmetry}, 
\emph{Phys. Rev.} {\bf D80}, 034508 (2009)

%F. D. R. Bonnet, R. G. Edwards, G. T. Fleming, R. Lewis, and D. G. Richards (LHP Collaboration), Phys. Rev. D 72, 054506 (2005).

\bibitem{RBC}
P. A. Boyle et al., \emph{The pion's electromagnetic form factor at small momentum transfer in full lattice QCD}, \emph{JHEP} {\bf 0807}, 112 (2008)

\bibitem{pacscs}
S. Aoki et al. (PACS-CS Collaboration), \emph{2+1 Flavor Lattice QCD toward the Physical Point}, \emph{Phys. Rev.} {\bf D79}, 034503 (2009)

%%%%%%%%%%%%%%%%%%%%%%%%%%%%

%\bibitem{bedaque2004}
%P.F. Bedaque, \emph{Aharonov-Bohm effect and nucleon nucleon phase shifts on the lattice}, \emph{Phys. Lett.}, {\bf B593}, 82 (2004)

%\bibitem{divitiis2004}
%G. M. de Divitiis, R. Petronzio, and N. Tantalo, \emph{On the discretization of physical momenta in lattice QCD }, \emph{Phys. Lett.}, {\bf B595}, 408 (2004)

\bibitem{boyle2007} 
Jonathan M. Flynn et al. \emph{Hadronic form factors in lattice QCD at small and vanishing momentum transfer, JHEP} {\bf 05}, 016 (2007)

\bibitem{sachrajda2005}
C. T. Sachrajda and G. Villadoro, \emph{Twisted boundary conditions in lattice simulations}, \emph{Phys. Lett.}, {\bf B609}, 73 (2005)

\bibitem{bedaque2005}
P.F. Bedaque and J.-W. Chen, \emph{Twisted valence quarks and hadron interactions on the lattice}, \emph{Phys. Lett.}, {\bf B616}, 208 (2005)

\bibitem{jian} F. J. Jiang and B. C. Tiburzi, \emph{Flavor twisted boundary conditions, pion momentum, and the pion electromagnetic 
form factor, Phys. Lett} {\bf B645}, 314 (2007)

%%%%%%%%%%%%%%%%%%%%%%%%%%%%

\bibitem{Z2_1}
K. Bitar et al, \emph{The QCD finite temperature transition and hybrid Monte Carlo, Nucl. Phys.} {\bf B313} 348, (1989)

\bibitem{Z2_2}
H.R. Fiebig and R.M. Woloshyn, \emph{Monopoles and chiral-symmetry breaking in three-dimensional lattice QED, Phys. Rev.} {\bf D42} 3520, (1990).

\bibitem{Z2_3}
S.-J. Dong and K.-F. Liu, \emph{Stochastic estimation with Z(2) noise}, \emph{Phys. Lett.} {\bf B328} (1994)
130 [{\tt arXiv:hep-lat/9308015}]

\bibitem{Z2_4}
M. Foster and C. Michael, \emph{Quark mass dependence of hadron masses from lattice QCD}, \emph{Phys. Rev.} {\bf D59} (1999) 074503

\bibitem{Z2_5}
C. McNeile and C. Michael, \emph{Decay width of light quark hybrid meson from the lattice}, {\emph Phys. Rev.} {\bf D73} (2006) 074506 [{\tt arXiv:hep-lat/0603007}]

\bibitem{OanhNguyen}
Oanh Hoang Nguyen, \emph{Pion form factor from 2+1 dynamical flavor lattice
QCD using the O(a) improved Wilson-clover quark
formalism}, \emph{PoS(LATTICE 2009)} 129 [{\tt arXiv:hep-lat/1003.3321}]

%ChPT analysis
\bibitem{RBC-chiral} RBC-UKQCD Collaborations, C. Allton {\it et al.}, \emph{Physical results from $2+1$ flavor domain-wall QCD and SU(2)  
chiral perturbation theory}, \emph{Phys. Rev. D} {\bf 78}, 114509 (2008).

% chPT
\bibitem{ChPT_NNLO}
J. Bijnens, G. Colangelo, and P. Talavera, \emph{The vector and scalar form factors of the pion to two loops} \emph{J. High Energy Phys.} {\bf 05} (1998) 014

\bibitem{ChPT_NNLO_3}
J. Bijnens, and P. Talavera, \emph{Pion and kaon electromagnetic form factors}, arXiv:hep-ph/0203049

\bibitem{ChPT_NNLO_4}
G. Colangelo, J. Gasser, and H. Leutwyler, \emph{$\pi\pi$ scattering}, \emph{Nucl. Phys.} {\bf B603}, 125 (2001).

\end{thebibliography}
\end{document}